\documentclass[fleqn,usenatbib]{mnras}

\usepackage{graphicx}	% Including figure files
\usepackage{amsmath}	% Advanced maths commands
\usepackage{amssymb}	% Extra maths symbols
\usepackage{fontawesome} % For github symbols etc
\usepackage{hyperref}
\usepackage[dvipsnames]{xcolor} % Extra colors
\usepackage{xspace}
\usepackage{orcidlink}
\usepackage{array}
\usepackage{xspace}
\usepackage{natbib}

\usepackage{todonotes} % for comments
\usepackage{siunitx}

\defcitealias{2018JCAP...07..041V}{\text{Paper} I}
\defcitealias{2020PhRvL.125k1101M}{\text{Paper} II}

\newcommand{\betahat}{\hat{\vect{\beta}}}

\newcommand{\muhat}{\hat{\vect{\mu}}}
\newcommand{\dd}{{\rm d}}
\newcommand{\vect}[1]{\boldsymbol{\mathbf{#1}}}

\definecolor{linkcolor}{rgb}{0.7752941176470588, 0.22078431372549023, 0.2262745098039215}

\definecolor{deepred}{rgb}{0.8,0.2,0.2}

\hypersetup{linkcolor=cyan,citecolor=magenta,filecolor=magenta,urlcolor=orange}

\newcommand{\lcdm}{$\Lambda \rm{CDM}$}
\newcommand{\gaia}{\textit{Gaia}\xspace}
\newcommand{\SNR}{\text{SNR}}
\newcommand{\AU}{\text{AU}}

\newcommand{\nbicon}{{\color{linkcolor}\faFileCodeO}\xspace}
\newcommand{\nblink}[1]{\href{https://github.com/kenvantilburg/lens-velocity/tree/main/#1}{\nbicon}}
\newcommand{\githubmaster}{\href{https://github.com/kenvantilburg/lens-velocity/}{\faGithub}\xspace}

\usepackage{newtxtext,newtxmath}
\usepackage[T1]{fontenc}

\DeclareRobustCommand{\VAN}[3]{#2}
\let\VANthebibliography\thebibliography
\def\thebibliography{\DeclareRobustCommand{\VAN}[3]{##3}\VANthebibliography}

\title[Astrometric Weak Lensing Constraints on DM substructure]{Astrometric Weak Lensing with \textit{Gaia} DR3 and Future Catalogs: Searches for Dark Matter Substructure}

\author[C. Mondino et al.]{{Cristina Mondino\orcidlink{0000-0002-8058-4055}},$^{1}$\thanks{cmondino@perimeterinstitute.ca}
	Andreas Tsantilas$^{2}$\thanks{andreas.tsantilas@nyu.edu } 
	Anna-Maria Taki\orcidlink{0000-0001-7976-2582},$^{3}$\thanks{annamariataki@gmail.com}
	Ken Van Tilburg\orcidlink{0000-0001-7085-6128},$^{2,4}$\thanks{kenvt@nyu.edu | kvantilburg@flatironinstitute.org} 
	Neal Weiner\orcidlink{0000-0003-2122-6511}$^{2}$\thanks{neal.weiner@nyu.edu}
	\\
	$^{1}$Perimeter Institute for Theoretical Physics, 31 Caroline St N, Waterloo, Ontario N2L 2Y5, Canada\\
	$^{2}$Center for Cosmology and Particle Physics, Department of Physics, New York University, 726 Broadway, New York, NY 10003, USA\\
	$^{3}$Institute for Fundamental Science, University of Oregon, 1371 E. 13th Ave, Eugene, OR 97403, USA\\
	$^{4}$ Center for Computational Astrophysics, Flatiron Institute, 162 5th Ave, New York, NY 10010, USA\\
}

\date{Accepted 2024 April 12. Received 2024 March 12; in original form 2023 September 7}

\pubyear{2023}

\begin{document}
	\label{firstpage}
	\pagerange{\pageref{firstpage}--\pageref{lastpage}}
	\maketitle
	
	\begin{abstract}
            Small-scale dark matter structures lighter than a billion solar masses are an important probe of primordial density fluctuations and dark matter microphysics. Due to their lack of starlight emission, their only guaranteed signatures are gravitational in nature.
            We report on results of a search for astrometric weak lensing by compact dark matter subhalos in the Milky Way with \textit{Gaia} DR3 data. Using a matched-filter analysis to look for correlated imprints of time-domain lensing on the proper motions of background stars in the Magellanic Clouds, we exclude order-unity substructure fractions in halos with masses $M_l$ between $10^7 \, M_\odot$ and $10^9 \, M_\odot$ and sizes of one parsec or smaller.
            We forecast that a similar approach based on proper accelerations across the entire sky with data from \textit{Gaia} DR4 may be sensitive to substructure fractions of $f_l \gtrsim 10^{-3}$ in the much lower mass range of $10 \, M_\odot \lesssim M_l \lesssim 3 \times 10^3 \, M_\odot$. We further propose an analogous technique for \emph{stacked} star-star lensing events in the regime of large impact parameters. Our first implementation is not yet sufficiently sensitive but serves as a useful diagnostic and calibration tool; future data releases should enable average stellar mass measurements using this stacking method. 
            \githubmaster
	\end{abstract}
	
	\begin{keywords}
		dark matter -- astrometry -- gravitational lensing
	\end{keywords}
	
	%%%%%%%%%%%%%%%%%%%%%%%%%%%%%%%%%%%%%%%%%%%%%%%%%%
	
	%%%%%%%%%%%%%%%%% BODY OF PAPER %%%%%%%%%%%%%%%%%%

\section{Introduction}\label{sec:intro}
The arrival of \gaia's second data release (DR2) heralded a new era in understanding the dynamics of the Milky Way (MW). The remarkable size and precision of this astrometric data set allowed scientists to pursue questions that had been previously unanswerable. As a result, we have learned of new stellar streams \cite{mateu2018fourteen,ibata2019streams,ibata2021charting,li2022s} and other galactic substructure \cite{belokurov2018co, helmi2018merger, koppelman2018one, myeong2018sausage, naidu2020evidence}.
The extremely accurate astrometry of \gaia's Early Data Release 3 (EDR3) has revealed the $5\ \mu\rm{as}/\rm{y}$ global pattern in the proper motions of extragalactic sources imprinted by the acceleration of the Solar System through aberration of light~\cite{2021A&A...649A...9G}, and has produced the most accurate construction of the celestial reference frame, based on 1.6 million quasar sources~\cite{klioner2022gaia}. Additionally, the enhanced quality of the EDR3 astrometric data has allowed for a comprehensive study of the distance and kinematic properties of the Magellanic Clouds~\cite{luri2021gaia} (geometrically anchoring the cosmic distance ladder), as well as for the exploration of the non-axisymmetric features of our own Galaxy~\cite{drimmel2022gaia}.

The idea of using gravitational microlensing to detect massive dark structures in the Galactic halo, as introduced in~\cite{1986ApJ...304....1P}, prompted several groups to combine astrometric and photometric measurements to search for dark compact halo objects~\cite{1995A&A...294..287H,1995AJ....110.1427M,1995ApJ...453...37W}.
Initially, photometric methods were suggested as the principal discovery channel for microlensing events~\cite{1991ApJ...371L..63P,1991ApJ...372L..79G, 1991ApJ...374L..37M}, with notable exceptions suggesting the use of astrometric data~\cite{dominik2000astrometric,Lu_2016}. The launch of the \gaia mission in 2013 signaled the dawn of precision astrometry and highlighted the prospects of astrometric probes in the hunt for compact dark halo objects~\cite{belokurov2002astrometric}.
The promise of astrometric measurements with unprecedented precision has since inspired the implementation of several astrometric lensing techniques on \textit{Gaia} data sets, aiming to identify the signatures of dark matter subhalos in the MW~\cite{erickcek2011astrometric, li2012new,2018JCAP...07..041V,2020PhRvL.125k1101M,vattis2021deep,mishra2022inferring,2023arXiv230100822C}. The analysis of astrometric microlensing data has led to the successful determination of the masses of nearby dwarf stars~\cite{2017Sci...356.1046S, 2018MNRAS.480..236Z, 2023MNRAS.520..259M}, and resulted in the first definitive detection of an isolated stellar-mass black hole~\cite{2022ApJS..260...55L,2022ApJ...933L..23L,2022ApJ...933...83S,2022ApJ...937L..24M}.

One exciting possibility with astrometric measurements of extraordinary precision is to explore the potential of \emph{time-domain astrometric lensing} in the study of gravitating bodies. As these objects move across the sky, they can induce a time-dependent shift in the positions of background objects, yielding potentially observable effects.
\cite{2018JCAP...07..041V}, hereafter \citetalias{2018JCAP...07..041V}, introduced different classes of observables to look for nonluminous structures in the Milky Way, solely via the time-varying astrometric imprints on background luminous sources.
Their detection strategies laid the groundwork for \textit{halometry}, creating a flexible framework to characterize the properties of a broad spectrum of DM substructures within the Galactic halo. 
Capitalizing on one of these techniques to tease out subtle time-domain astrometric lensing effects, \cite{2020PhRvL.125k1101M}, hereafter \citetalias{2020PhRvL.125k1101M}, performed the first search for individual DM subhalos in the MW, by implementing a matched-filter template of local lensing distortions to the proper motions of Magellanic Cloud stars in the \gaia DR2 data. This proof-of-concept analysis constrained parsec-sized (or smaller) lenses in the $10^7$--$10^8 \, M_\odot$ mass range.
Although this technique exhibited sensitivity to more dilute subhalos, which could have otherwise  escaped detection with photometric lensing probes, the results are currently statistics-limited. With increased observational periods and resolved systematics, the sensitivity of the search is expected to improve steeply over time, with the capability to deliver parametric leaps in reach with data from ongoing and planned astrometric surveys.  

In this paper, we perform an improved search using proper motion templates for lensing by DM subhalos on \gaia DR3 data following~\citetalias{2020PhRvL.125k1101M}, with the expected sensitivity enhancements borne out in this data release.  We also outline extensions of this technique to parallax and acceleration templates; forecasts for the latter show exceptional promise for lighter subhalos, in the range $10 \, M_\odot$--$10^6 \, M_\odot$.  Since acceleration measurements are currently absent from the \gaia archive, we demonstrate a method to extract them by combining data from the DR2/DR3 releases; we will perform an acceleration template analysis on this catalog in follow-up work. While the proper motion template analysis presented in this paper will eventually hit a noise floor from intrinsic stellar motion dispersion -- at least on the Magellanic Clouds (MCs) -- the acceleration analysis is expected to be statistics-limited for the foreseeable future, with even steeper improvements expected after longer observation times.

We furthermore suggest that a \emph{stacked} version of our template methods can look for \emph{collective} star-star lensing in the astrometric, nontransient regime. To this end, we construct a catalog of optical doubles of foreground and background stars at substantially different line-of-sight distances, but at (accidentally) small angular separations. We execute the first star-star astrometric lensing analysis based on proper motion templates, and do not (yet) find evidence for this effect, in line with expectations. In the process, we identify several potential biases and systematics in \gaia's astrometric data for stellar pairs at very close separations. The inclusion of stellar accelerations in a future analysis of this type should yield a positive detection by \gaia DR4, potentially establishing an estimator for the average mass of the foreground stellar population, and serving as a calibrator for the DM searches.

This paper is organized as follows. In section~\ref{sec:dm_lensing}, we investigate the full potential of searches for DM substructure using template observables. We give a review of their definitions in section~\ref{sec:dm_templ}, estimate current and future sensitivities in section~\ref{sec:vel_acc_proj}, and present an updated limit from \gaia DR3 data in section~\ref{sec:veltempl_update}. In section~\ref{sec:star-star}, we describe the new application of template observables to measure astrometric lensing in a collection of optical doubles. We present a test statistic that combines the three template observables in section~\ref{sec:combined_tau}, detail the sample selection and data processing in section~\ref{sec:star-star_data}, and summarize our findings in section~\ref{sec:star-star_res}. We conclude in section~\ref{sec:conclusion}. Additionally, we discuss possible improvements and limitations of our test statistic for dark subhalo searches in appendix~\ref{app1:multi-lens}, we compute for the first time the lensing-induced parallax and define the corresponding template observable in appendix~\ref{app2:parallax}, and outline the extraction of accelerations from DR2/DR3 data with an estimate of their uncertainty in appendix~\ref{app2:acceleration}.

We publicly release our data analysis pipeline for the DM search and the star-star lensing analysis, allowing their use by other groups and on other astrometric data sets (not necessarily from \gaia). The code used to obtain the results of this study is available on GitHub \githubmaster and a link below each figure (\nbicon) provides the Python code with which it was generated. Associated data are publicly available at \href{https://users.flatironinstitute.org/~kvantilburg/lens-velocity/}{this link}.

\section{Dark Matter lensing}\label{sec:dm_lensing}
Modern cosmology has cemented the role that Dark Matter (DM) plays in shaping the properties and growth of structures in the Universe. While the particle nature of DM is unknown, it is determined to be non-relativistic from $z\lesssim 10^7$ and onward, and to have scale-invariant adiabatic perturbations, which grow under the influence of gravity. These perturbations eventually form halos that merge hierarchically, yielding a nearly scale-invariant spectrum of structures with masses $M_\text{halo} \gtrsim 10^9 \, M_\odot$. Clustering on smaller scales is largely unconstrained nonetheless. A variety of mechanisms -- increased primordial perturbations, phase transitions, attractive interactions or dissipative dynamics -- can lead to enhanced clustering at smaller scales far beyond what is predicted in \lcdm. The main observational challenge is to detect these low-mass subhalos, which are entirely dark due to their lack of star formation or even baryonic content, indirectly through their gravitational interactions. Proposed detection methods (see \cite{2022arXiv220307354B} for a recent overview) include strong lensing~\cite{2010MNRAS.408.1969V,2016ApJ...823...37H,2017JCAP...05..037B}, stellar streams~\cite{2019ApJ...880...38B}, lensing of gravitational waves~\cite{2018PhRvD..98j4029D}, pulsar timing arrays~\cite{2019PhRvD.100b3003D,2020JCAP...12..033R}, and time-domain astrometric weak lensing~\citepalias{2018JCAP...07..041V}. In this work, we focus on the last method.

As dark subhalos move across the celestial sphere, the apparent locations of background luminous objects will undergo time-dependent shifts due to weak gravitational lensing. Specifically, the proper motion, acceleration, and parallax of the foreground subhalo will imprint \emph{correlated} lensing-induced shifts to the proper motion, acceleration, and parallax of any background objects in the angular vicinity of the lens. 
While the lensing-induced motion of any individual background source is impossible to detect in the large-impact-parameter regime (at small impact parameters, searches for astrometric transients can be powerful~\cite{2023arXiv230100822C}), the collective, correlated lensing-induced motions have a characteristic pattern that can be teased out statistically on \emph{multiple} sources through a ``template'' analysis. These template observables are matched filters to the lensing effects on the background linear motion, linear acceleration, and parallax. 

Previous works have considered microlensing effects over multiple background sources. For example, \cite{2005ApJ...635..711G} investigated the possibility of detecting outer solar system objects using astrometric microlensing induced by the lens parallax motion, while \cite{2008ApJ...684...46D} suggested to look for nearby and fast-moving lenses through a combination of photometric and astrometric lensing effects on multiple stars. Both works considered lenses that span large angles on the sky during the observation time, while here we focus on the regime of small relative changes in the lens-source impact parameters, where ``template'' observables can be defined.

\subsection{Template observables}\label{sec:dm_templ}

Template observables are designed to search for local, correlated astrometric weak lensing effects induced by extended dark lenses on multiple luminous sources.  
Proper motion and angular acceleration templates were first introduced in \citetalias{2018JCAP...07..041V} and an optimized version of the proper motion template was used in \citetalias{2020PhRvL.125k1101M} to search for DM-induced signals on Magellanic Clouds stars from \gaia's second data release (DR2). Since template observables are the main focus of this work, we summarize here the relevant existing results.

In the weak lensing regime, the lens-source angular separation is much larger than the Einstein radius of the lens and one image of the source is completely resolved. In this case,    
%In the weak lensing regime, 
a gravitational lens $l$ with mass $M_l$ which appears to be near a background luminous source $i$ induces an apparent angular deflection to the source's true position on the plane perpendicular to the line of sight,
\begin{align}
    \label{eq:ang_defl}
    \Delta\vect{ \theta}_i = -\frac{4 G M_l}{c^2 r_l} \frac{\widetilde{M}_l(\beta_{li})}{\beta_{li}/\beta_l} \hat{\vect{\beta}}_{li},
\end{align}
where $G$ is Newton's gravitational constant, $c$ the speed of light, and $r_l$ the physical size of the lens. We also define the angular size $\beta_l \equiv r_l / D_l$ of the lens. The angular separation (impact parameter) between the source and the lens is denoted as  $\vect{\beta}_{li} \equiv \vect{\theta}_l - \vect{\theta}_i$. The function $\widetilde{M}_l(\beta_{li}) \equiv \widetilde{M}_l(\beta_{li})/M_l$ is the normalized lens mass enclosed within the impact parameter, and we assume the source distance to be much larger than the lens distance $D_l$. 

In general, the impact parameter changes over time due to the relative motion of the lens, the observer, and the source -- $\vect{\beta}_{li} = \vect{\beta}_{li}(t)$ -- leading to a time-varying lensing angular deflection, which results into an apparent motion of the background source. When the relative change in impact parameter over the total observation time is small, $|\Delta \vect{\beta}_{li}| \ll |\vect{\beta}_{li}|$, the time-dependent lensing correction can be extracted through a series expansion of Eq.~\eqref{eq:ang_defl}. The dominant contribution to the time-dependent impact parameter comes from the linear proper motion $\vect{\mu}_l$ of the lens with 3D physical velocity $\vect{v}$ and line-of-sight direction $\hat{\vect{D}}_l$, parametrized as:
\begin{align}
    \vect{\beta}_{li}(t) = \vect{\beta}^0_{li} + \vect{\mu}_l t, \qquad \vect{\mu}_l \equiv \frac{v_{l \perp}}{D_l} [\hat{\vect{v}}_l - \hat{\vect{D}}_l (\hat{\vect{D}}_l \cdot \hat{\vect{v}}_l)] \equiv \frac{\vect{v}_{l \perp}}{D_l},
    \label{eq:lens_proper_motion}
\end{align}
with $t = 0$ corresponding to the time of closest approach, at angular impact parameter $\vect{\beta}_{li} = \vect{\beta}_{li}^0$.
For $\mu_l \tau \ll \beta^0_{li}$ over the observation time $\tau$, the leading-order effect is then a lens-induced proper motion
\begin{equation}
    \Delta\vect{ \mu}_{li} =  \frac{4 G M_l \mu_l}{c^2\beta_l^2 D_l} \widetilde{ \vect{\mu}}(\beta_l, \vect{\beta}_{li}, \hat{\vect{\mu}}_l), 
    \label{eq:lens_pm}
\end{equation}
with a dipole-like profile:
\begin{equation}
    \label{eq:lens_pm_prof}
    \widetilde{ \vect{\mu}} (\beta_l, \vect{\beta}, \muhat_l) \equiv \frac{\widetilde{M}_l(\beta)}{\beta^2 / \beta_l^2} \left[ 2\betahat (\betahat \cdot \muhat_l) - \muhat_l \right]  - \frac{\partial_{\beta} \widetilde{M}_l(\beta)}{\beta/ \beta_l^2} \betahat (\betahat \cdot \muhat_l).
\end{equation}
The next-to-leading order correction is a lens-induced angular acceleration
\begin{equation}
    \label{eq:lens_acc}
    \Delta\vect{ \alpha}_{li}  = \frac{8 G M_l \mu_l^2}{c^2 \beta_l^3 D_l} \widetilde{\vect{\alpha}}(\beta_l, \vect{\beta}_{li}, \muhat_l).
\end{equation}
with a quadrupole-like profile
\begin{align}
    \label{eq:lens_acc_prof}
    \widetilde{\vect{\alpha}}(\beta_l, \vect{\beta}, \muhat_l ) 
    &= \frac{\widetilde{M}_l(\beta)}{\beta^3 / \beta_l^3} \left[ 2 \muhat_l \left( \betahat \cdot \muhat_l \right) + \betahat \left( 1- 4(\betahat \cdot \muhat_l )^2 \right)  \right]  \nonumber \\
    &\phantom{=} - \frac{\partial_{\beta } \widetilde{M}_l(\beta )}{\beta^2 /\beta_l^3} \left[ \muhat_l (\betahat \cdot \muhat_l ) +  \frac{1}{2} \betahat \left( 1 - 5 (\betahat \cdot \muhat_l)^2 \right) \right]  \nonumber \\
    &\phantom{=} - \frac{\partial^2_{\beta } \widetilde{M}_l(\beta )}{\beta /\beta_l^3} \frac{1}{2} \betahat (\betahat \cdot \muhat_l )^2.
\end{align}
For a given lens mass distribution $\widetilde{M}$, the two-dimensional spatial profiles of the time-domain lensing distortions, $\widetilde{\vect{\mu}}$ and $\widetilde{\vect{\alpha}}$, depend only on the position $\vect{\beta}$, the characteristic angular scale $\beta_l$, and the velocity direction $\muhat_l$. 
    
The universal nature of the lensing distortion patterns from Eqs.~\eqref{eq:lens_pm_prof} and \eqref{eq:lens_acc_prof} allows us to define the template observables $\mathcal{T}_\mu$ and $\mathcal{T}_\alpha$, which quantify the overlap between the measured proper motions $\lbrace \vect{\mu}_i \rbrace$ and accelerations $\lbrace \vect{\alpha}_i \rbrace$, respectively, of a field of stars with the \emph{expected} distribution in the presence of a lensing signal. For a candidate lens at location $\vect{\theta}_t$, with angular size $\beta_t$, and velocity direction $\muhat_t$, we define
\begin{align}
    \label{eq:template}
    \mathcal{T}_\mu(\beta_t,  \vect{\theta}_t, \muhat_t) \equiv \sum_i \frac{\vect{\mu}_i \cdot \widetilde{\vect{\mu}}(\beta_t,  \vect{\theta}_t - \vect{\theta}_i, \muhat_t)}{\sigma_{\mu, i}^2} \\
    \mathcal{N}^2_\mu(\beta_t,  \vect{\theta}_t, \muhat_t) \equiv \sum_i \frac{ |\widetilde{\vect{\mu}}(\beta_t,  \vect{\theta}_t - \vect{\theta}_i, \muhat_t)|^2}{\sigma_{\mu, i}^2},
\end{align}
for stars at locations $\lbrace \vect{\theta}_i \rbrace$ in the surroundings of $\vect{\theta}_t$ with measured proper motion variance $\sigma_{\mu, i}$. We also introduce the convenient normalization factor $\mathcal{N}_\mu$. Analogous statistics $\mathcal{T}_\alpha$ and $ \mathcal{N}_\alpha$ can be defined for lensing-induced accelerations.

If the mean motions of the background sources are zero (or subtracted) and the errors are Gaussian and uncorrelated, maximizing $\mathcal{T}$ is equivalent to maximizing the log-likelihood ratio of the hypothesis of a local signal (i.e.~the presence of a lens in a given location with corresponding angular scale, and velocity direction) versus that of the null hypothesis. When searching for a population of dark lenses in front of a stellar target, the lens properties are usually unknown and need to be marginalized over, requiring a refined version of the $\mathcal{T}$ test statistic, such as the global test statistic $\mathcal{R}$ introduced in \citetalias{2020PhRvL.125k1101M}, utilized in the DR3 analysis in Sec.~\ref{sec:veltempl_update}, and discussed in appendix~\ref{app1:multi-lens}. Nevertheless, the local template from Eq.~\eqref{eq:template} still captures the relevant features and its simple expression provides an easy analytic estimate of the local signal-to-noise ratio (SNR) and thus the best stellar targets in the search for a lensing signal from a dark subhalo. 
    
By construction, the lens-induced angular acceleration in Eq.~\eqref{eq:lens_acc} is a subleading contribution to the lensing correction, and one could naively think that the proper motion signal always dominates. However, the sensitivity of each observable strongly depends on the properties of the measured luminous sources and is ultimately limited by the intrinsic variance of the stellar motions. We will compare the sensitivity of proper motion and acceleration templates to compact dark lenses and show their complementarity to cover different regions of the parameter space in section~\ref{sec:vel_acc_proj}.

Other types of lensing corrections can arise from a lens motion which is different from the linear motion considered so far. For example, the source-lens impact parameter also changes over time due to parallax, as $\vect{\beta}_{li}(t) \simeq \varpi_l \left( \cos \omega t, \sin \delta_\text{ecl} \sin \omega t \right)$, where $\varpi_l$ is the lens parallax, $\omega \simeq 2\pi/\text{y}$ is the orbital angular velocity, and $\delta_\text{ecl}$ is the ecliptic latitude of the lens -- neglecting the background source's parallax and assuming for simplicity that the observer moves in a circular orbit. Expanding the angular shift in Eq.~\eqref{eq:ang_defl} for $\varpi_l \ll \beta^0_{li}$, the lens-induced parallax of the luminous source can be obtained and the corresponding template observable can be defined. The derivation of this anomalous parallax contribution and estimates for the sensitivity of the parallax template observable, compared to the proper motion template, are presented in appendix~\ref{app2:parallax}. When the astrometric measurements are statistically limited, $\mathcal{T}_{\mu}$ is typically more sensitive than $\mathcal{T}_{\varpi}$, as the total distance travelled by a dark lens with velocity $v_l \sim 10^{-3}c$ is usually much larger than the parallax displacement of one AU, especially over multi-year observations. However, when the proper motion dispersion is limited by the intrinsic stellar motion, $\mathcal{T}_{\mu}$ and $\mathcal{T}_{\varpi}$ can have comparable sensitivities (see appendix~\ref{app2:parallax}). For other types of lenses, such as outer Solar System planets, which move slowly with respect to the observer, the anomalous parallax is the dominant effect~\cite{2005ApJ...635..711G}; \citetalias{2018JCAP...07..041V}.
    
We conclude this section by highlighting the difference between the lensing regime in which the template approximation is valid and where a different treatment is needed. As pointed out earlier, template observables correctly capture the astrometric lensing corrections when the arc on the sky spanned by the lens motion during the survey is smaller than the impact parameter $\beta_{li}$. For lenses with a finite characteristic size $\beta_l = r_l/D_l$, the SNR for the lensing-induced proper motion is maximized at $\beta_{li} \sim \beta_l$, unless the lens has a very cuspy inner profile (see Sec.~2 of \citetalias{2018JCAP...07..041V}) or is effectively point-like, i.e.~when $\beta_l$ is smaller than the typical star-star separation, which for \gaia is $\gtrsim 0.7''$ \cite{2021A&A...649A...5F}. Therefore, the template searches in this work are designed for lenses with $r_l \gtrsim 0.003\ \textrm{pc}\ (v_{l \perp}/10^{-3}c)(\tau/10\ \textrm{y})$ for the proper motion and acceleration effects, and $r_l > \AU$ for anomalous parallax. When this condition is not satisfied, e.g.~for very compact, cuspy, or fast-moving lenses, the astrometric lensing effect manifests as a transient perturbation to the trajectory of individual (or multiple) stars that can be searched for using other observables, referred to as mono-blips (or multi-blips) in \citetalias{2018JCAP...07..041V} -- see \cite{2023arXiv230100822C} for a projection of the mono-blip observable in (mock) \gaia DR4 data. \\

\subsection{Sensitivity estimates and projections} 
\label{sec:vel_acc_proj}

Proper motion and acceleration template searches in current and upcoming \gaia data can be sensitive to signals from Galactic compact dark lenses, thus probing parts of previously unexplored parameter space of dark matter substructure. For the sensitivity forecasts, we briefly review here the estimates of the local SNR for the template observables and refer the reader to Sec.~4.2 of \citetalias{2018JCAP...07..041V} for a more extensive derivation. 
    
The observed stellar population is assumed to have zero or subtracted background motion and uncorrelated Gaussian noise, such that $\langle \mathcal{T} \rangle_\text{noise} =0 $ and $\langle {\mathcal{T}}^2 \rangle_\text{noise} = {\mathcal{N}}^2$. 
On a field of background stars with angular number density $\Sigma_0$, one expects that $ {\mathcal{N}}^2 \sim \Sigma_0 \beta_l^2/{\sigma^2_{\mu, \alpha}}$ up to an $\mathcal{O}(1)$ numerical factor that depends on the characteristic lens density profile. For a template that perfectly matches the true lens properties, $\langle \mathcal{T} \rangle_\text{signal} = C_{\mu, \alpha} {\mathcal{N}}^2$, where $C_\mu = 4 G M_l \mu_l / c^2 \beta_l^2 D_l$ and $C_\alpha = 8 G M_l \mu_l^2 / c^2 \beta_l^3 D_l$. Therefore, the SNR for the template observable is
    \begin{align} \label{eq:snr}
    	\SNR_{\mathcal{T}}  =  \frac{C_{\mu, \alpha} \sqrt{\Sigma_0} \beta_l}{\sigma_{\mu,\alpha}} = 
        	\begin{cases}
         		 \frac{4 G M_l \mu_l }{c^2 \beta_l D_l } \frac{ \sqrt{\Sigma_0}}{\sigma_\mu}  \ \ \ \ \ \ \ \ \ \ \ \ \ \ \  \ \ \ \textrm{for}\ \mu, \\
         		 \frac{8 G M_l \mu_l^2 }{c^2 \beta_l^2 D_l } \frac{ \sqrt{\Sigma_0}}{\sigma_\alpha}   \ \ \ \ \ \ \ \ \ \ \ \ \ \ \  \ \ \ \textrm{for}\  \alpha. 
       	 \end{cases}\,
    \end{align}
This expression makes evident the fact that the proper motion and acceleration observables are the two leading terms in the Taylor expansion in $\mu_l/\beta_l = v_{l\perp}/r_l$ of the time-domain lensing signal.
The SNR in Eq.~\eqref{eq:snr} is dominated by the most massive, fastest, nearby, and compact lens in the parameter combinations shown. 

The variation in $v_{l\perp} = \mu_l D_l$ is independent of the other lens properties and given by the DM virial velocity dispersion in the Galaxy $\langle v_{l \perp} \rangle = \sqrt{\pi/2}\sigma_{v, \rm{DM}}$, with $\sigma_{v, \rm{DM}} \simeq 166\ \rm{km}/\rm{s}$. 
We assume the lens population is distributed according to the Galactic density profile $\rho_l = f_l \rho_\mathrm{DM}$, where $f_l$ is the (constant) fraction in DM substructure in halos of mass $M_l$. The closest lens in front of a stellar target with angular number density $\Sigma_0$ is expected at a distance of
    \begin{align} \label{eq:Dl_min}
    	D_{l, \text{min}} \simeq \left( \frac{3 M_l}{\Delta \Omega \, \rho_l} \right)^{1/3},
    \end{align}
where $\rho_\mathrm{DM}$ (and thus $\rho_l$) is approximated to be constant up to $D_{l, \text{min}}$.
The smallest lens size $r_{l,\text{min}}$ allowed in the template regime is fixed by the lens displacement over the observational time $v_{l\perp}  \tau  = \mu_l D_l \tau$ (see section~\ref{sec:dm_templ}). However, the sensitivity will not improve further for lenses that are effectively point-like, i.e.~when their angular size is smaller than the typical angular separation $\sim 1/\sqrt{\pi \Sigma_0}$ of background sources. Therefore,
    \begin{align} \label{eq:rl_min}
    	r_{l, \text{min}} = \text{max} \left\lbrace  v_{l\perp} \tau, \frac{1}{\sqrt{\pi \Sigma_0}} D_{l, \rm{min}} \right\rbrace.
    \end{align}
    
    \begin{figure} 
        \includegraphics[width=\columnwidth]{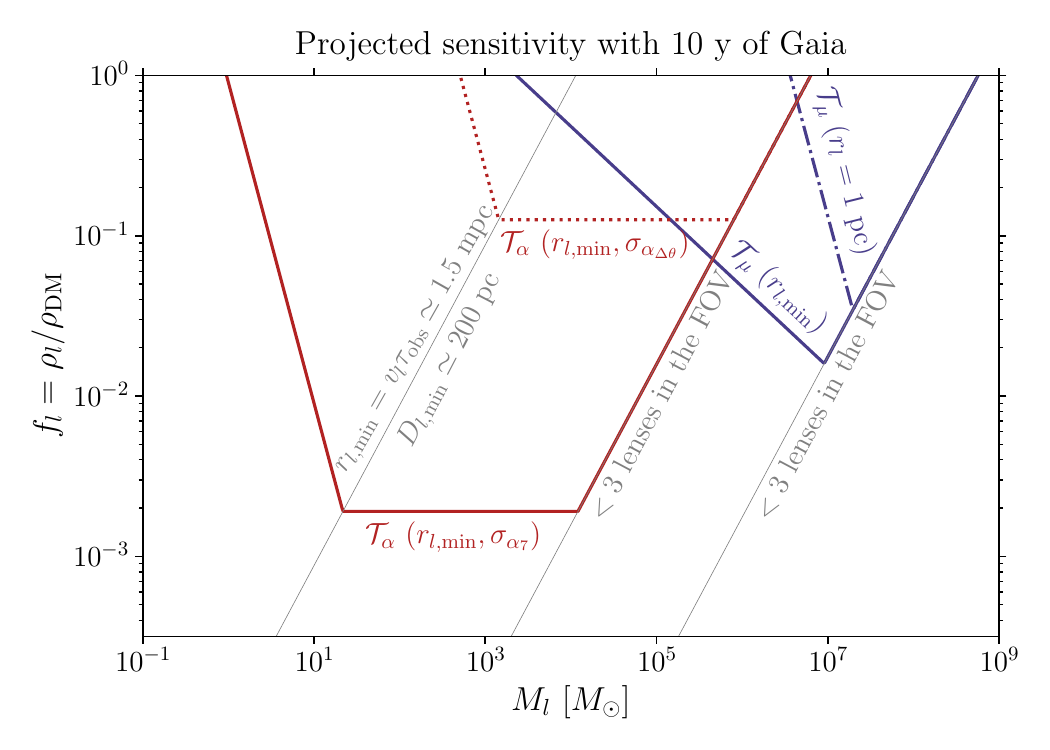}
        \caption{Sensitivity projections for proper motion and angular acceleration templates in the parameter space of lens fractional abundance $f_l  = \rho_l/\rho_\text{DM}$ versus mass $M_l$. All curves correspond to local $\SNR_{\mathcal{T}}=1$ (optimistically ignoring the look-elsewhere effect), with the signal-to-noise ratio given in Eq.~\eqref{eq:snr}, taking $v_{l \perp} = 208\ \text{km}/\text{s}$ and the smallest lens distance from Eq.~\eqref{eq:Dl_min}. The proper motion template (blue contours) is evaluated for the minimum lens size given in Eq.~\eqref{eq:rl_min} (solid) and a fixed value of $r_l = 1$ pc (dash-dotted), using the LMC as stellar target ($\Sigma_0 = 5\times 10^8\ \text{rad}^{-2}, \Delta\Omega=0.02\ \text{rad}^2, \sigma_\mu = 0.2$ mas/y, $D_\text{stars} = 50$ kpc). The acceleration template (red contours) is evaluated for the minimum lens size using Galactic Disk stars ($\Sigma_0 = 4.5\times 10^9\ \text{rad}^{-2}, \Delta\Omega=0.2\ \text{rad}^2$, $D_\text{stars} = 3$ kpc), taking $\sigma_{\alpha_7} = 4.5\ \mu\text{as}/\text{y}^2$ (solid) and $\sigma_{\alpha_{\Delta \theta}} = 300\ \mu\text{as}/\text{y}^2$(dotted) -- assuming per-epoch positional accuracy of $200\ \mu \rm{as}$. \nblink{dm_limit/modules/plot_limit.ipynb}}
        \label{fig:template_projections}
    \end{figure}
    
In Fig.~\ref{fig:template_projections} we show ${\rm{SNR}}=1$ sensitivity projections for proper motion and angular acceleration templates for 10 years of \gaia observations, using the above values of $D_{l, \text{min}}$ and $r_{l, \text{min}} $ in Eq.~\eqref{eq:snr}. The sensitivity depends on the proper motion and acceleration dispersions, $\sigma_\mu$ and $\sigma_\alpha$, of the observed stellar populations. If these were only instrumentally limited, $N_{\rm{obs}}$ observations with individual position uncertainty $\sigma_{\delta\theta}$ would result in $\sigma_\mu \sim \sigma_{\delta\theta}/(\tau \sqrt{N_{\rm{obs}}})$ and $\sigma_\alpha \sim \sigma_{\delta\theta}/(\tau^2 \sqrt{N_{\rm{obs}}})$; we estimate the numerical factors of this parametric dependence in appendix~\ref{app2:acceleration}. In this case, ${\rm{SNR}}_{\mathcal{T}_\alpha}/{\rm{SNR}}_{\mathcal{T}_\mu}=v_{l\perp} \tau/r_l < 1$, i.e.~the proper motion template would always perform better, as expected, since it is the leading-order effect when $v_{l\perp} \tau < r_l$. 
The instrumental precision for most of the stars observed by \gaia is already below the intrinsic proper motion dispersion, or will be so by the end of the mission. %On the other hand, a measurement of angular accelerations by \gaia would only be statistically limited because the intrinsic accelerations -- from the Galactic potential, wide binary companions, or exoplanets -- are typically far below the survey's sensitivity. 
On the other hand, a measurement of angular accelerations by Gaia would only be statistically limited because the intrinsic accelerations — from the Galactic potential, wide binary companions, or exoplanets — are typically far below the survey’s sensitivity on an individual distant star, and are uncorrelated among nearby stars (a relatively small fraction of nearby and/or bright stars do have detectable intrinsic accelerations from binary companions~\cite{2023A&A...674A...9H, 2022MNRAS.513.5270P, 2018A&A...614A..30R}).
Consequently, we see in Fig.~\ref{fig:template_projections} that acceleration templates offer the best prospects for measuring a compact DM subhalo lensing signal throughout a large portion of the parameter space, particularly for masses below $\sim 10^{6}\ M_\odot$. 

For each observable, we consider the best stellar target, i.e.~the one that maximizes the figure of merit $\sqrt{\Sigma_0}/\sigma_{\mu, \alpha}$. The best performance is obtained by a $\mathcal{T}_\mu$ search on the Large Magellanic Cloud (LMC) and a $\mathcal{T}_\alpha$ search on Galactic Disk stars. An additional advantage of using angular accelerations is the possibility of using a larger star sample, spanning a wider portion of the sky, while minimizing intrinsic proper motion dispersion requires choosing stars that are further away (but still in a dense field).
The SNR scaling obtained above breaks down at large lens masses, when Poisson fluctuations of the number of lenses in the field of view (FOV) become important. For our sensitivity projections of Fig.~\ref{fig:template_projections}, we require more than 3 lenses in front of the stellar target.
    
Currently, angular accelerations for sources observed by \gaia are not directly available. The optimal way to measure the acceleration is to include it as an additional parameter to the astrometric fit. However, before the full time series of \gaia's observations become available with DR4, an indirect measurement of 
accelerations can be obtained by combining position and proper motion measurements at different times from the DR2 and DR3 catalogs. In appendix~\ref{app2:acceleration}, we outline a method to construct such an acceleration catalog together with an estimate of the expected statistical uncertainty. The implementation of this method and a careful investigation of the systematics involved is left to future work~\cite{acc_paper}. Applying a template search for DM lensing effects on this catalog would already produce interesting results, as shown in Fig.~\ref{fig:template_projections}, where we refer to the expected uncertainty on the derived acceleration as $\sigma_{\alpha_{\Delta \theta}}$, while $\sigma_{\alpha_7}$ refers to the statistically-limited uncertainty from a 7-parameter astrometric fit. The possible applications of an acceleration catalog for \gaia's sources of course go beyond DM searches with astrometric weak lensing and include measuring collective star-star lensing effects (see section~\ref{sec:star-star}), mapping out the Milky Way potential \cite{Buschmann:2021izy}, and searching for ultra-low frequency gravitational waves~\cite{1996ApJ...465..566P,2011PhRvD..83b4024B,2018CQGra..35d5005K}.

\subsection{Search for Dark Matter Substructure with \gaia DR3} \label{sec:veltempl_update}

The proper motion template observable described in the previous sections has already been successfully applied to \gaia DR2 data to search for lensing signals induced by galactic DM subhalos on MCs stars in \citetalias{2020PhRvL.125k1101M}. Here we repeat the same analysis with the improved astrometry of \gaia DR3 \cite{2021A&A...649A...2L,luri2021gaia}. We closely follow the procedure described in \citetalias{2020PhRvL.125k1101M}; we only briefly review it here and refer the reader to the original paper for more details.

Firstly, the \gaia archive is queried for stars with G magnitude measurements and parallaxes consistent with zero at $5\sigma$. By implementing the following selections, we choose stars that are located within a $5^\circ$ radius from the center of the Large Magellanic Cloud (LMC),
\\ \\ \texttt{SELECT *  FROM gaiaedr3.gaia\_source\\ WHERE 1=CONTAINS(POINT('ICRS',ra,dec), \\ CIRCLE('ICRS',81.28,-69.78,5))  AND parallax/parallax\_error < 5 AND\\ phot\_g\_mean\_mag > 0} \\ \\
and within a $4^\circ$ radius from the center of the Small Magellanic Cloud (SMC),\\
\\ \texttt{SELECT * FROM gaiaedr3.gaia\_source\\ WHERE 1=CONTAINS(POINT('ICRS',ra,dec),\\CIRCLE('ICRS',12.80,-73.15,4)) AND parallax/parallax\_error < 5 AND\\ phot\_g\_mean\_mag > 0.}\\ \\
To remove the globular clusters NGC 104 (47 Tuc) and NGC 362 from the SMC's foreground, we impose additional cuts on the stellar proper motions, so that $|\mu_{\alpha*} - 0.685 \text{ mas/y}| < 2 \text{ mas/y}$ and $|\mu_{\delta} + 1.230 \text{ mas/y}| < 2 \text{ mas/y}$.

Secondly, the data are {\it cleaned} in two steps:
\begin{enumerate}
    \item \label{itm:first_clean} The background mean proper motion field computed with a Gaussian distance kernel of radius $0.1^\circ$ is subtracted from the stars' proper motion; the smoothed angular number density map is computed with a Gaussian distance kernel of radius $0.1^\circ$ and overdense pixels of about $0.014^\circ$ are removed if their density is $2.5$ times larger than the average to reduce contamination from star clusters; proper motion outliers at more than $5\sigma$ are removed and the following additional cuts on the parallax and the quality of the astrometric fit are imposed \cite{2021A&A...649A...2L}: \texttt{parallax/parallax\_error < 2, ruwe < 1.4, ipd\_gof\_harmonic\_amplitude < 0.4, ipd\_frac\_multi\_peak < 40, ipd\_frac\_odd\_win < 40}. \\
    \item \label{itm:second_clean} The background mean proper motion field computed with a Gaussian distance kernel of radius $0.06^\circ$ is subtracted and proper motion outliers at more than $3\sigma$ are removed, repeating the procedure 3 times; the effective proper motion dispersion as a function of the stars' G magnitude and distance from the center of each cloud are computed using G magnitude bins of size $0.1$ and radial bins of size $1^\circ$ to group the stars.
\end{enumerate}

The optimized global test statistic $\mathcal{R}$ from Eq.~(9) in \citetalias{2020PhRvL.125k1101M}, obtained by maximizing the likelihood ratio of the signal hypothesis over background, is evaluated on the end products of the procedure described in~\ref{itm:second_clean}. The same \textit{clean} data are also used as input for the data-driven noise + signal simulations as described in \citetalias{2020PhRvL.125k1101M}. The processing step in~\ref{itm:second_clean} is performed on the simulated data, before the evaluation of the test statistic. For each simulation, we fix the lens population parameters $\lbrace M_l, r_l, f_l \equiv \rho_l/\rho_\text{DM} \rbrace$, where $\rho_\text{DM}$ is the Milky Way DM halo density profile, which we model as an NFW profile $\rho_\mathrm{DM} =  4 \rho_s / [(r/r_s)(1+r/r_s)^2]$ with scale radius $r_s = 18\ \rm{kpc}$ and density at the scale radius of $\rho_s = 0.003\  M_{\odot}/\rm{pc}^3$.\footnote{This treatment is conservative, as the DM density towards the LMC may be enhanced by up to a factor of 2 due to tidal gravitational interactions \cite{2019ApJ...884...51G, 2021ApJ...919..109G}. Furthermore, the LMC will likely have carried its own subhalos in the MW, further increasing dark substructure. Hence, upper limits on $f_l$ that are larger than unity (but less than 2 or 3) in our conservative model can be physical.} The distribution of the test statistic from 150 simulations per point is then compared to the observed value in the data and the parameter space point is excluded at $90\% (50\%)$ CL if $\mathcal{R}_\text{data} (M_l, r_l, f_l ) <  \mathcal{R}^{\text{90\% (50\%) }} _{\text{sim}}(M_l, r_l, f_l )$. 

\begin{figure} 
    \includegraphics[width=\columnwidth]{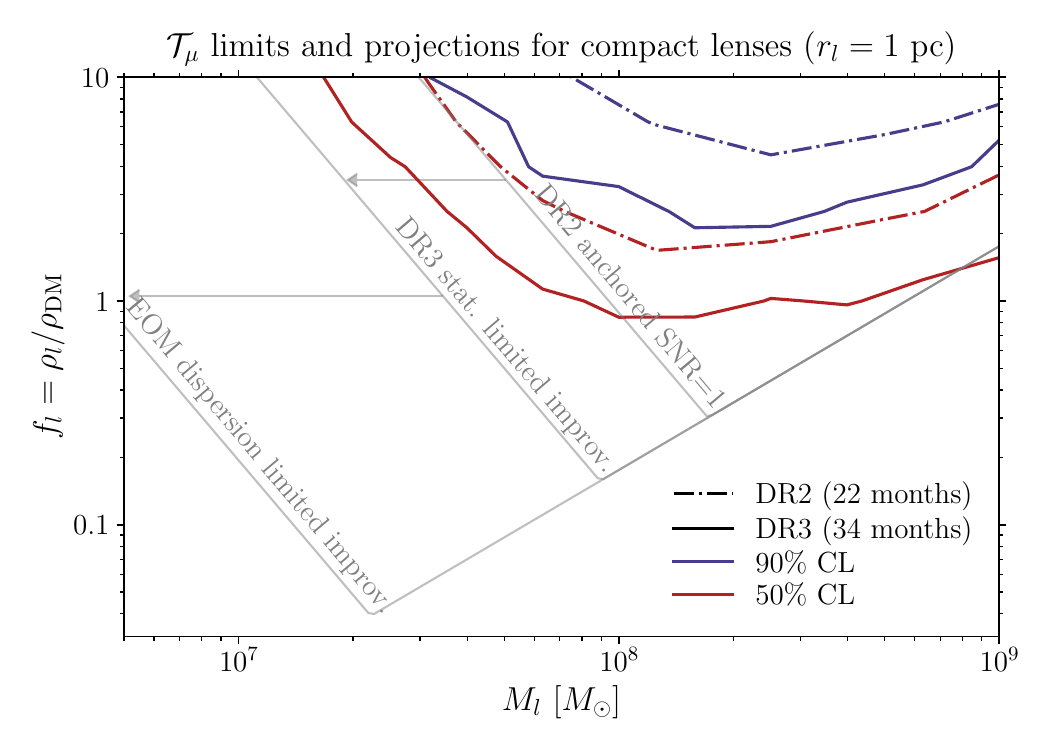}
    \caption{Limits and projected sensitivity on the fraction of dark compact lenses $f_l$ with mass $M_l$ and characteristic size $r_l = 1$ pc from the proper motion template analysis on MC stars. The solid blue (red) line shows the $90\%\ (50\%)$ CL limit from the analysis on \gaia DR3 data, while  the dash-dotted lines correspond to the previously obtained results from \protect\citetalias{2020PhRvL.125k1101M}. In light gray we show the $\SNR_{\mathcal{T}_\mu}=1$ curves, where $\SNR_{\mathcal{T}_\mu}$ is given in Eq.~\eqref{eq:snr}, $D_l = D_{l, \text{min}}$ from Eq.~\eqref{eq:Dl_min}, $r_l = 1$ pc, $\Sigma_0 = 5\times 10^8\ \text{rad}^{-2}$, $\Delta\Omega=0.02\ \text{rad}^2$, and $v_{l \perp} = 208\ \text{km}/\text{s}$. The proper motion dispersion $\sigma_\mu$ is fixed to $1.75$ mas/y to anchor the $\SNR_{\mathcal{T}_\mu}=1$ line  to the $50\%$ CL DR2 limit, scaled by a factor of $0.5$ to obtain the statistics-limited optimistic DR3 improvement (see Eq.~\eqref{eq:sigma_mu_improv}), and fixed to an intrinsic dispersion of $0.2$ mas/y for the EOM projections. Differently from Fig.~\ref{fig:template_projections}, here an estimate of the look-elsewhere effect is included in the analytic projections by anchoring the DR2 line to the observed limit and taking into account the LMC average effective dispersion of $1.54$ mas/y in DR2 (see Fig.~\ref{fig:pm_disp}) \nblink{dm_limit/modules/plot_limit.ipynb}}
    \label{fig:limit_compact}
\end{figure}

\begin{figure} 
    \includegraphics[width=\columnwidth]{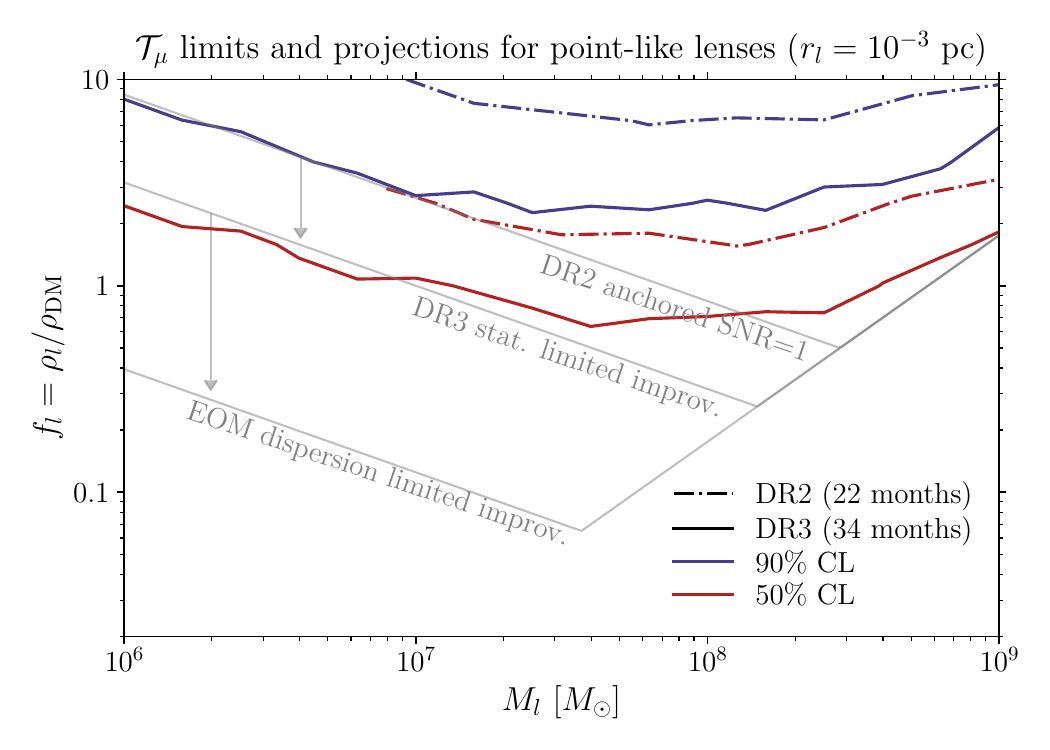}
    \caption{Same as Fig.~\ref{fig:limit_compact}, but for point-like lenses with $r_l = 10^{-3}\ \mathrm{pc}$. The gray $\SNR_{\mathcal{T}_\mu}=1$ curves here correspond to $r_l = r_{l, \text{min}}$ from Eq.~\eqref{eq:rl_min} and the proper motion dispersion $\sigma_\mu$ is fixed to $6.3$ mas/y to anchor the $\SNR_{\mathcal{T}_\mu}=1$ line  to the $50\%$ CL DR2 limit. \nblink{dm_limit/modules/plot_limit.ipynb}  }	
    \label{fig:limit_pointlike}
\end{figure}

\begin{figure}
    \includegraphics[width=\columnwidth]{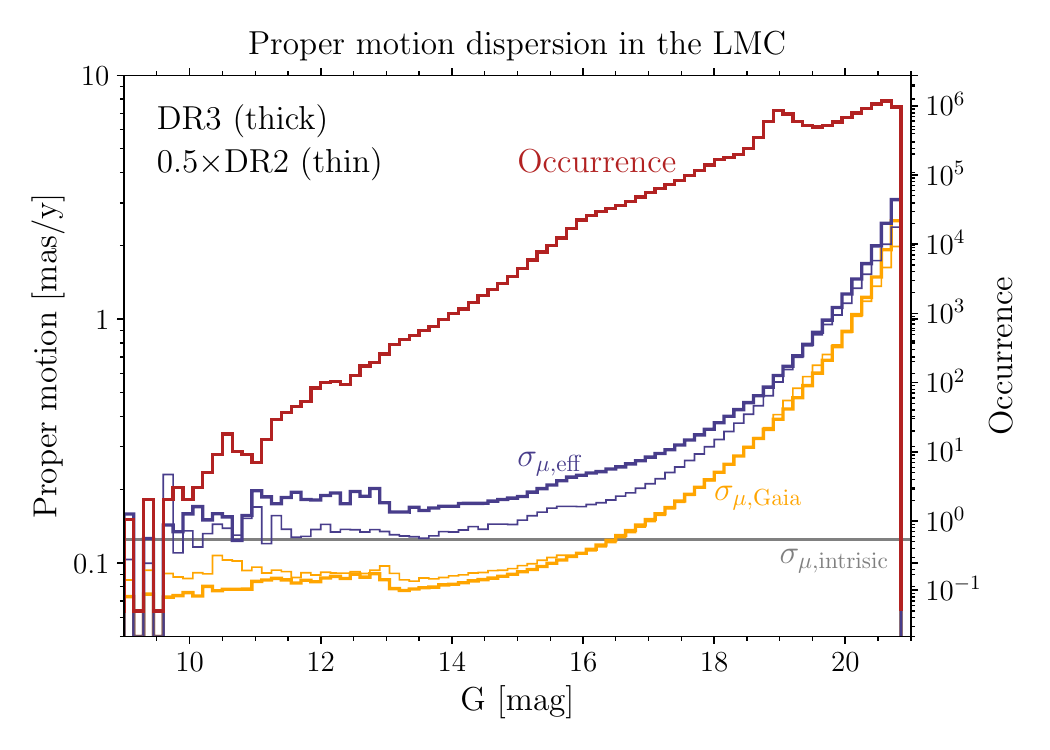}
    \caption{Number of stars (red), effective (blue), reported (yellow), and intrinsic (gray) proper motion dispersion as a function of stellar G magnitude for the LMC. The thick lines correspond to \gaia DR3 data, while the thin lines for $\sigma_{\mu, \text{eff}}$ and $\sigma_{\mu, \text{Gaia}}$ correspond to \gaia DR2 data rescaled by a factor of 0.5 that corresponds to the expected statistical improvement with time (see Eq.~\eqref{eq:sigma_mu_improv}). 
    \nblink{dm_limit/modules/data_cleaning.ipynb} }
    \label{fig:pm_disp}
\end{figure}

Following the procedure described above, we can exclude the presence of dark lenses in front of the MCs. The results are shown in the parameter space of fractional lens abundance $f_l $ versus lens mass $M_l$, for compact lenses ($r_l = 1\ \text{pc}$) in Fig.~\ref{fig:limit_compact} and point-like lenses ($r_l = 10^{-3}\ \text{pc}$) in Fig.~\ref{fig:limit_pointlike}. 
We compare the updated limits obtained with the analysis on \gaia DR3 stars with the ones obtained with the analysis on \gaia DR2. The improvement in the constraints comes from the expected improvement on the stellar proper motion measurements due to the longer observation time, 
\begin{align} \label{eq:sigma_mu_improv}
	\sigma_{\mu}^{\text{DR3}} \simeq \left(\frac{22\ \rm{months}}{34\ \rm{months}}\right)^{3/2} \sigma_{\mu}^\text{DR2} \simeq 0.5\ \sigma_{\mu}^\text{DR2}.
\end{align}
As shown in Fig.~\ref{fig:pm_disp}, this improvement is borne out in \gaia's reported error $\sigma_{\mu, \text{Gaia}} \equiv \langle \sigma^2_{\mu, i} \rangle ^{1/2}$ for all the stars in the LMC. However, the relevant quantity for the proper motion template evaluation is the effective proper motion dispersion $\sigma_{\mu, \text{eff}} \equiv \langle (\mu_{i} - \mu_\text{mean})^2 \rangle ^{1/2}$ observed in the stellar population, which improves by a factor of 2 only for stars with $19 \lesssim G \lesssim 20$. The dispersion at lower and higher G magnitude values appears not to be statistics-limited and therefore has a somewhat less favorable scaling with time. 

In Figs.~\ref{fig:limit_compact} and~\ref{fig:limit_pointlike}, the observed DR2 and DR3 limits on dark lenses are compared with the optimistic improvement expected for statistics-limited proper motion variance, using the SNR scaling derived in section~\ref{sec:vel_acc_proj}. 
The expected statistic-limited improvement on the fractional abundance of the dark lenses is 
\begin{align} \label{eq:ratio}
    f_l & \propto  
    \begin{cases}
        \sigma_\mu^3 \propto t^{-9/2}       \qquad \qquad & \text{fixed } r_l ~(\text{Fig.}~\ref{fig:limit_compact}) \\
        \sigma_\mu^{3/2} \propto t^{-9/4}   \qquad & r_{l, \text{min}} ~(\text{Fig.}~\ref{fig:limit_pointlike})
   \end{cases}\,.
\end{align}
The projected sensitivity for \gaia's end of the mission (EOM) results is also shown, assuming that the effective dispersion will be completely dominated by the MCs' intrinsic proper motion $\sigma_{\mu, \text{eff}} = \sigma_{\mu, \text{intrinsic}} \simeq 0.2$ mas/y.

The global template test statistic $\mathcal{R}$ used in the analysis above relies on the strongest signal produced by an individual lens in front of the stellar target. In appendix~\ref{app1:multi-lens} we discuss a possible generalization that includes multiple lenses and show how it is not expected to improve over current results -- with our conservative treatment for noise modeling -- because the signal decreases faster than the noise for subleading lenses (due to the look-elsewhere effect). Further investigation of an optimal test statistic that leverages on multiple lenses while also not being hurt by strong look-elsewhere effects go beyond the scope of this work.

Several other gravitational probes of massive compact objects which make up all or a fraction of the DM abundance have been investigated in the literature. The most well-studied candidates are primordial black holes (PBHs), which are strongly constrained in the mass range $10^6$--$10^9\ \mathrm{M}_{\odot}$ (see for example~\cite{Carr:2020gox, Bird:2022wvk}). Some of the existing bounds that would rule out all the parameter space of Fig.~3 for \emph{point-like lense} only apply specifically to PBHs; these are effects that rely on accretion, such as CMB spectral distortions and anisotropies~\cite{Ricotti:2007au, Ali-Haimoud:2016mbv} and direct X-ray observations~\cite{Inoue:2017csr}. Dynamical effects on the MW~\cite{1994ApJ...437..184X}, globular clusters~\cite{1993ApJ...413L..93M, Brandt:2016aco}, dwarf galaxies~\cite{Lu:2020bmd, Takhistov:2021aqx}\footnote{Notice that typically the dominant PBHs gas heating mechanisms in dwarf galaxies are related to accretion, and not dynamical friction.}, and wide binaries~\cite{Yoo:2003fr, 2009MNRAS.396L..11Q} would probably also be induced by the most compact dark lenses considered here; however in order to derive robust constraints from these effects, dedicated galactic simulations and direct comparison to data are required to obtain reliable constraints from dynamical friction and heating effects induced by compact (but finite size) subhalos~\cite{Arvanitaki:2019rax}. PBHs constraints based on the induced formation of cosmic structure incompatible with LSS~\cite{Carr:2018rid} and Lyman-$\alpha$ forest observations~\cite{Murgia:2019duy} are model dependent, as they are sensitive to the cosmic history and formation time of the subhalos. Finally, traditional photometric microlensing surveys rule out ultracompact lenses with masses below about $10^3\ M_{\odot}$~\cite{Blaineau:2022nhy}, which are lighter than the ones considered here, and furthermore require much smaller subhalo scale radii (smaller than their Einstein radii) for the strong-lensing signal to be unsuppressed.

\section{Star-Star lensing} \label{sec:star-star}

The previous section expounded on a blind search for astrometric lensing effects from dark subhalos, lenses whose properties are, at best, only known at the population level. 
We now consider a different kind of lens target: isolated stars in the \gaia catalog that, by chance, happen to be angularly close to other, more distant stars. We will refer to such resolvable optical doubles as pairs of foreground and background stars. In this case, most properties of the lens -- its location, velocity, and distance -- are known, except its mass. The measurement of lensing corrections induced in the motion of the background stars can therefore be used for directly inferring the mass of the foreground (lens) star.
The possibility of a mass measurement for individual stellar lenses through astrometric microlensing has been investigated in \cite{1998ApJ...502..538B,2001A&A...375..701D,2017ApJ...843..145K,2018MNRAS.476.2013R}.  
Different predictions for the expected number of close encounters that would produce a microlensing event measurable by \gaia have been forecast by \cite{kluter2018prediction, mcgill2020predictions,2022AJ....164..253L} and a precision better than~$15\%$ in the mass measurement is expected for a dozen events \cite{kluter2020expectations}. 

We suggest a complementary approach, which aims at detecting \emph{collective} weak lensing distortions for a large number of stellar pairs at wider angular separations, where the lensing event is \emph{not} a transient on typical survey time scales. The cumulative effect from multiple star-star lensing corrections can be captured using our template observables and searched for in existing and upcoming astrometric data sets, including those of \gaia.

To gain intuition on the sensitivity of current and future data, we can estimate the SNR within some approximations. The signal is dominated by optical doubles at the smallest observable angular separation $\beta_{\rm{min}}$, and with the lens at the smallest possible distance $D_{l, \rm{min}}$. However, the template regime only applies for $D_{l, \rm{min}} \beta_{\rm{min}} > v_{l \perp}\tau$. The best-case scenario is thus at the saturation of this lower bound. We can \emph{stack} all of the star-star lensing events and compute collective test statistics $\mathcal{T}^\star_{\mu}$ and $\mathcal{T}^\star_\alpha$ to arrive at SNRs analogous to those in Eq.~\eqref{eq:snr}:
\begin{align} \label{eq:snr_ss_mu}
    \SNR_{\mathcal{T}_{\mu}^{\star}}  & = 
        \frac{8\pi G M_\odot \sqrt{v_{l \perp} \Sigma_0 n_{l}}}{\sigma_\mu \sqrt{\tau\beta_{\rm{min}}} } \nonumber \\
        & \simeq 1.3\ \frac{3\ \rm{mas}/\rm{y}}{\sigma_{\mu}}\sqrt{\frac{v_{l \perp}}{10^{-3}c}\frac{\Sigma_0}{8\times 10^7}\frac{n_l}{0.1\ \rm{pc}^{-3}}\frac{0.7''}{\beta_{\rm{min}}}\frac{34\ \rm{months}}{\tau} }, \\
        \label{eq:snr_ss_alpha}
    \SNR_{\mathcal{T}_{\alpha}^{\star}}  & = 
        \frac{8\pi G M_\odot \sqrt{2 v_{l \perp} \Sigma_0 n_{l}}}{\sigma_\alpha \tau \sqrt{3\tau\beta_{\rm{min}}} }  \nonumber \\
        & \simeq 1.1\ \frac{\rm{mas}/\rm{y}^2}{\sigma_{\alpha}} \sqrt{\frac{v_{l \perp}}{10^{-3}c}\frac{\Sigma_0}{8\times 10^7}\frac{n_l}{0.1\ \rm{pc}^{-3}} \frac{0.7''}{\beta_{\rm{min}}} \left(\frac{34\ \rm{months}}{\tau}\right)^{3}} ,
\end{align}
for the proper motion and the acceleration templates, respectively. In the above, we have assumed solar-mass lenses with a constant number density $n_l$ up to $D_{l, \rm{min}}$ -- about a few hundred pc -- and a uniform angular number density $\Sigma_0$ of distant background stars. The smallest star-star separation is limited by \gaia's completeness to be $\beta_{\rm min} \sim 0.7''$~\cite{2021A&A...649A...5F}. 

From Eq.~\eqref{eq:snr_ss_mu}, we see that a star-star lensing detection using proper motion templates on \gaia DR3 data is challenging. If the proper motion dispersion of background stars were statistically limited, it would scale as $\sigma_{\mu} \propto \tau^{-3/2}$ and the sensitivity would improve linearly with time. However, from the analysis presented in the next sections, we find that $\sigma_{\mu}$ is most likely already limited by the intrinsic stellar motion and will not decrease with time. On the other hand, a near-future detection using angular acceleration templates looks promising. In Eq.~\eqref{eq:snr_ss_alpha}, we used the dispersion expected for accelerations inferred from the combination of \gaia DR2 and DR3 observations (see appendix~\ref{app2:acceleration}). However, a dedicated 7-parameter astrometric fit including angular accelerations could lead to:
\begin{align}
    & \sigma_{\alpha} \simeq 25\ \mu\rm{as}/\rm{y}^2 \left( \frac{66\ \rm{months}}{\tau} \right)^{5/2}, \\
    & \SNR_{\mathcal{T}_{\alpha}^{\star}} \simeq 17\ \frac{\tau}{66\ \rm{months}}, \label{eq:snr_ss_alpha_2}
\end{align}
where we used a per-epoch positional accuracy of $250\ \mu \rm{as}$, reflecting \gaia's astrometric performances on stars with magnitude $G=17$; for fainter stars with $G=19$, the sensitivity worsens by about a factor of 3.6. Effectively, we predict that \gaia DR4 should produce a high-SNR measurement of collective star-star lensing. Given the scaling with observation time, our approach may lead to meaningful mass measurements at the population level (for isolated stars) by the end of the \gaia mission ($\tau \sim 10\ \rm{y}$) if instrumental systematics are kept at bay.

In the following, we first describe how template observables can be easily adapted to search for star-star time-domain gravitational lensing signals. We then present an analysis applying proper motion templates on a sample of foreground-background stars selected from the \gaia DR3 catalog. As expected, a detection is not yet possible, but our implementation of the analysis pipeline provides insights on the sensitivity to the signal beyond the estimates of Eq.~\eqref{eq:snr_ss_mu}, identifies parts of phase space with systematic effects, and sets the stage for a future detection using stellar accelerations. The technique would enable a measurement of a weighted \emph{population average} mass for the foreground stars, instead of individual masses. With sufficient sensitivity, the star-lenses could be grouped in classes of stars with similar properties, as inferred from photometric and spectroscopic information, to obtain average mass measurements of more homogeneous populations. Finally, template analyses of star-star lensing serve as a valuable calibration for searches of dark lenses that rely on similar techniques (section~\ref{sec:dm_lensing}).

\subsection{Combined template observables}  \label{sec:combined_tau}

Consider a pair of accidentally-close, but resolvable, stars on the sky at different line-of-sight distances: the foreground star, $l$, will act as a gravitational lens on the background star, $s$, inducing the same time-domain astrometric lensing effects as described in section~\ref{sec:dm_templ}. In this case, the lens is point-like, with $\widetilde{M}_l(\beta) = 1$ in Eqs.~\eqref{eq:ang_defl}-\eqref{eq:lens_acc_prof}, and known distance, position and velocity. For each foreground-background pair, the signal is too small to be detected individually, but we can add up the contributions from many such pairs $p$ with angular separation $\beta_{ls} < \beta_{\rm{max}}$, where the maximum separation taken to be $\beta_{\rm{max}} = 3''$\footnote{This choice is arbitrary but of little consequence, since the signal decreases quadratically with increasing impact parameter.}. We define the \textit{star-star} proper motion template observable and its normalization as:
\begin{align} \label{eq:tau_mu_starstar}
    \mathcal{T}^{\rm{\star}}_\mu & = \sum_p (\vect{\mu}^p_{s})^{T} \cdot (\Sigma^p_{\mu, s})^{-1} \cdot \Delta \vect{\mu}^p_{ls} , \nonumber \\
    {\mathcal{N}^{\rm{\star}}_\mu} & = \sqrt{\sum_p \left(\Delta \vect{\mu}^p_{ls}\right)^T \cdot \left(\Sigma^p_{\mu, s}\right)^{-1} \cdot \Delta \vect{\mu}^p_{ls}},
\end{align}
where $\vect{\mu}^p_{s}$ and $\Sigma^p_{\mu, s}$ are the proper motion and proper motion covariance matrix of the background star in the pair $p$. Notice that a single foreground star could be lensing multiple background stars and so appear in multiple pairs in Eq.~\eqref{eq:tau_mu_starstar}. We define the lensing distortion as the one produced by a lens of one solar mass,
\begin{align} \label{eq:mu_starstar}
     \Delta \vect{\mu}_{ls} & =  \left(1-\frac{D_l}{D_s}\right) \frac{4 G M_\odot \mu_{ls}}{c^2 D_{l}} \frac{2 \hat{\vect{\beta}}_{ls}  (\hat{\vect{\beta}}_{ls} \cdot  \hat{\vect{\mu}}_{ls} ) - \hat{\vect{\mu}}_{ls} }{\beta_{ls}^2},
\end{align}
where $\vect{\beta}_{ls} = \vect{\theta}_l - \vect{\theta}_s$, $\vect{\mu}_{ls} = \vect{\mu}_l - \vect{\mu}_s$, and $D_{s (l)}$ denotes the distance to the source (lens). 
Similar template observables can be defined for the acceleration and parallax, $\mathcal{T}^{\star}_\alpha$ and $\mathcal{T}^{\star}_\varpi$. The three templates can be combined in a single test statistic
\begin{align} \label{eq:combined_tau}
	\mathcal{T}^{\star}_\text{all} & =  \mathcal{T}^{\star}_\mu + \mathcal{T}^{\star}_\varpi + \mathcal{T}^{\star}_\alpha \\
	\mathcal{N}^{\star}_\text{all} & = \sqrt{ (\mathcal{N}^{\star}_\mu)^2 + (\mathcal{N}^{\star}_\varpi)^{2} + (\mathcal{N}^{\star}_\alpha)^2}.
\end{align}
The expected signal-to-noise ratio for the star-star lensing signal is $\SNR_{\mathcal{T}^\star} = \langle m_l \rangle \mathcal{N}^{\star}_\text{all}$, where $\langle m_l \rangle$ is an {\it average} lens mass in solar mass units. An estimator for the average stellar mass is thus:
\begin{align} \label{eq:combined_tau}
	\langle m_l \rangle = \frac{ \mathcal{T}^{\star}_\text{all}  }{ (\mathcal{N}^{\star}_\text{all} )^2 }.
\end{align}
If an estimator of the foreground star mass is available, i.e.~as estimated from photometry or spectroscopy, it can be included as an additional weight in the lens-induced distortion of Eqs.~\eqref{eq:mu_starstar}\footnote{\gaia DR3 uses BP/RP spectra to derive astrophysical parameters, including stellar masses, for 140 million sources \cite{2023A&A...674A..26C}. We discuss the the effect of including such measurements on our template observable in section~\ref{sec:star_astro_param}.}. 

The \gaia catalog contains several millions of accidentally-close foreground-background pairs as defined in this work. 
Below, we describe our selection of the data sample and present an analysis with the proper motion template observable. We do not include the angular acceleration, due to the lack of acceleration measurements for now, nor the parallax template, due to the challenges in subtracting the background (see section~\ref{sec:star-star_res} for further details). Once a catalog of angular accelerations is built, it will be straightforward to extend our analysis to include the acceleration template.

\subsection{Data sample and background subtraction}  \label{sec:star-star_data}

To compute the above star-star lensing template, we create a catalog of \emph{optical doubles} within \gaia DR3\footnote{The analysis was performed on stars from EDR3, but the results would not change on DR3 data, since the astrometric solution is unchanged.}. Any of these accidental doubles consists of a foreground star -- the lens $l$ -- and a background star -- the source $s$. Specifically, we searched for all stellar pairs satisfying the two conditions $|\vect{\beta}_{ls}| < \beta_\mathrm{max}$ and $\varpi_l - \varpi_s > n_{\sigma} \sqrt{\sigma_{\varpi_l}^2 + \sigma_{\varpi_s}^2}$. In other words, we selected stars within $\beta_\text{max} = 3''$ from each other on the celestial sphere, but at significantly different line-of-sight distance (at significance of $n_\sigma = 2$), thereby reducing the contamination from binary stars and incorrectly-tagged pairs.

\begin{figure*}   
    \includegraphics[width=0.49\textwidth]{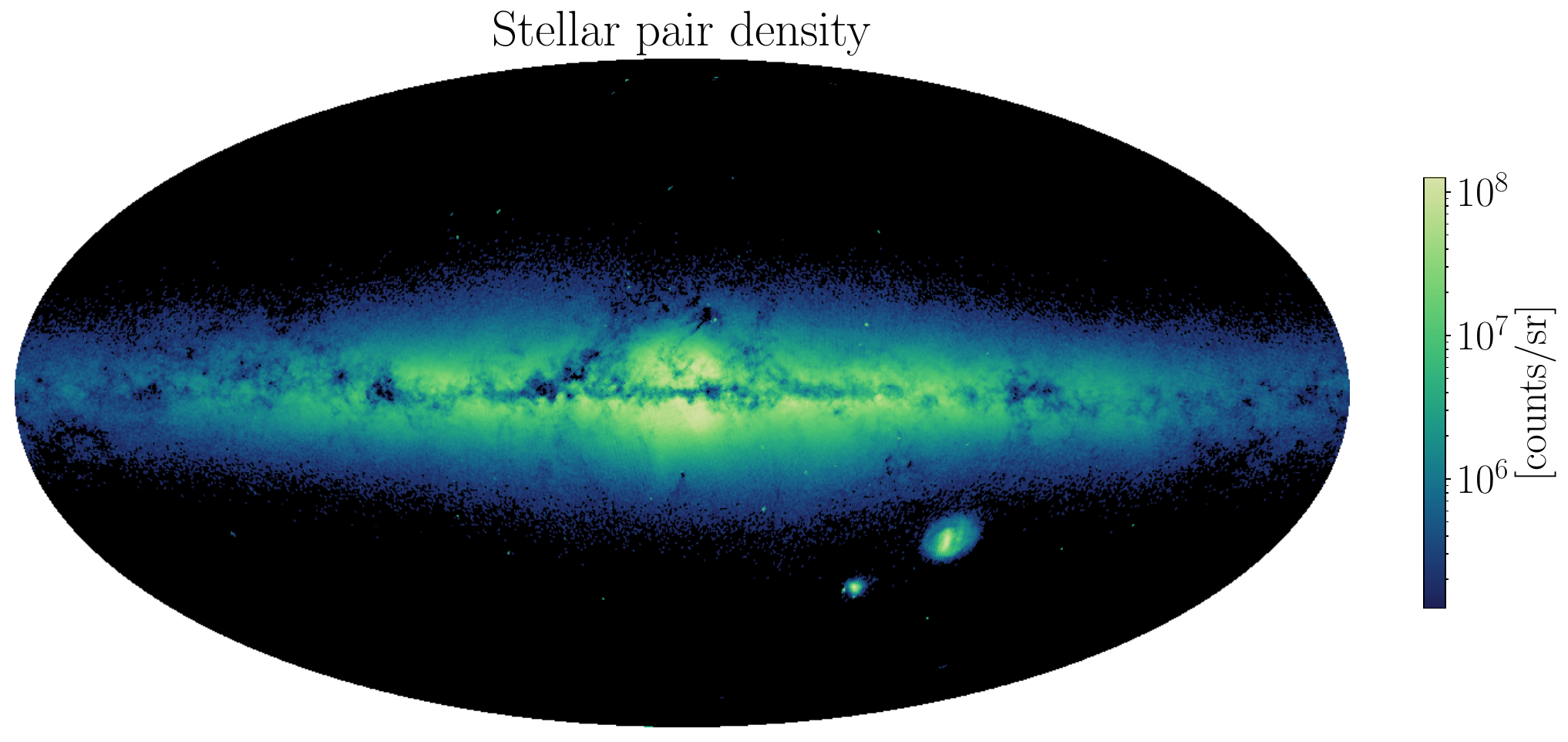}
    \includegraphics[width=0.49\textwidth]{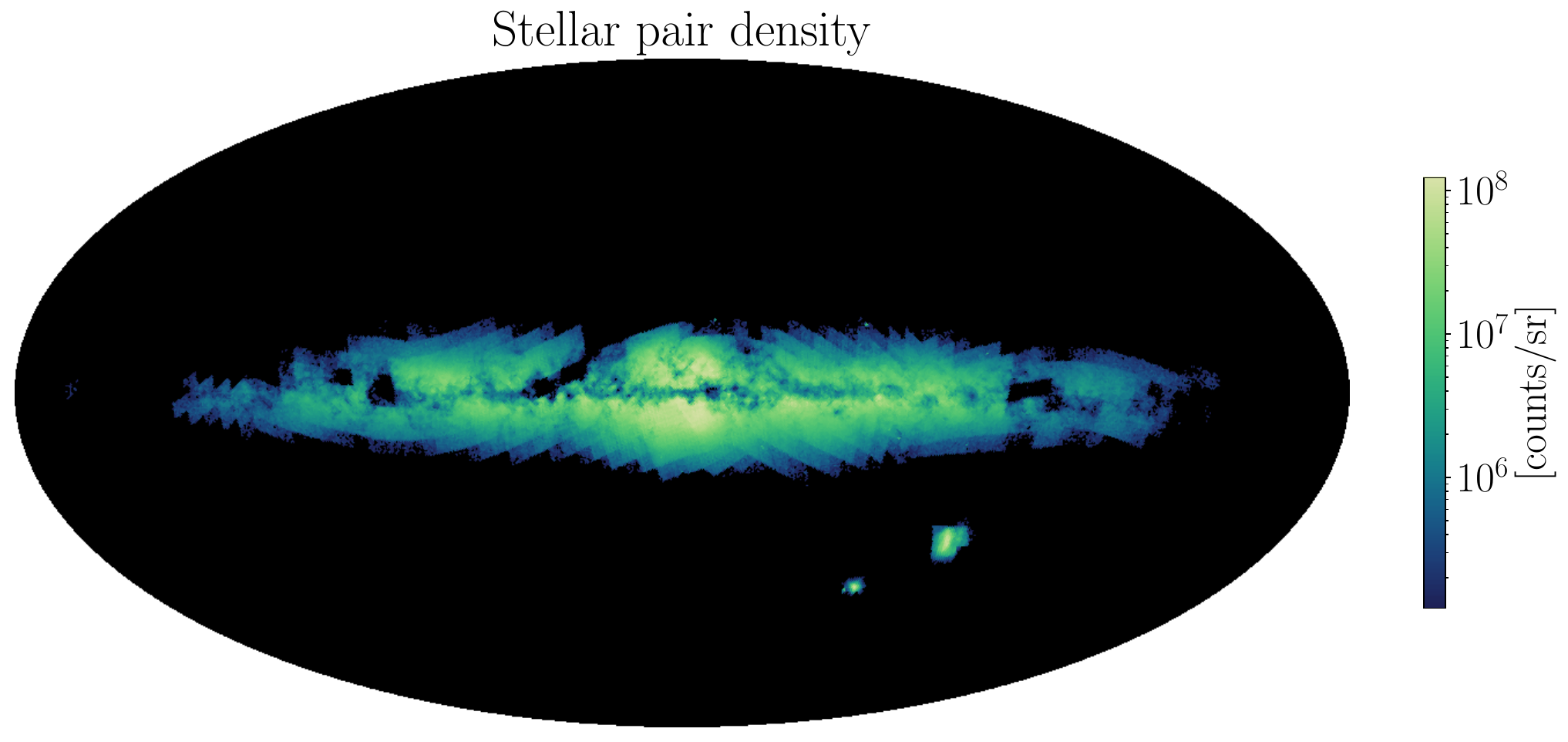}
    \caption{Angular number density of foreground stars in the sample of {\it optical doubles} selected from \gaia's DR3 on a pixel size of about $0.45\ \rm{deg}$ in Galactic coordinates. The visible large-scale structure is caused by extinction dropping the background source density in the optical wavelengths. ({\it left}) Sample of 15.1 million pairs after the cuts on $\texttt{RUWE}$ and stellar distances described in section~\ref{sec:quality_distance}. ({\it right}) Sample of 11.7 million pairs after the background subtraction and proper motion outlier removal described in section~\ref{sec:ss_back_sub}; the visible spatial features are due to the spatial  HEALPix pixelation on a scale of about $ 3.7\ \rm{deg}$ during the background subtraction and subsequent removal of stars falling in ``sparse'' bins (see the main text for further details). \nblink{star-star/make_plots.ipynb}}	
    \label{fig:doubles_map}
\end{figure*}

With those criteria, we identified about 61 million optical doubles. 
In the following sections, we describe a procedure whereby we further select stellar pairs within the original sample to ensure quality of the astrometric fit, reduce contamination from incorrectly-classified pairs, subtract the mean proper motion from the population of background stars, and remove proper motion outliers. After applying these additional selection cuts, we retain a clean sample of about 11.4 million optical doubles that will be used to evaluate the lensing template. A sky density map of the selected stellar pairs is shown in Fig.~\ref{fig:doubles_map}.

\subsubsection{Quality and distance cuts} \label{sec:quality_distance}
To ensure a high quality of the astrometric data used in our analysis, we require the goodness-of-fit statistic $\texttt{RUWE}$ to be $< 1.4$ \cite{2021A&A...649A...2L}. Furthermore, since for most of \gaia's stars the inverse parallax is a poor estimate of the distance due to large fractional uncertainties, we make use of the probabilistic stellar distances estimated in \cite{2021AJ....161..147B} to further ensure that foreground sources are indeed in front of their background counterparts -- in addition to the parallax selection described above. \cite{2021AJ....161..147B} provides two types of distances: geometric, combining parallax together with a direction-dependent
prior on distance, and photogeometric, which additionally uses color and apparent magnitude measurements. We use their geometric distances, since accidental doubles occur in dense regions of the sky so their color might be incorrectly estimated. 
We impose the condition $d_{s} - d_{l} > n_\sigma \sqrt{\sigma^2_{l,\text{86th}}+ \sigma ^2_{s,\text{14th}}}$, where $d_{l(s)}$ is the posterior median of the geometric distance of the lens (source), and $\sigma_{l,\text{86th}}$ and $\sigma_{s,\text{14th}}$ denote the difference between the posterior median and the 86th and 14th percentiles for the lens and the source, respectively. The quality and distance cuts heavily reduce the size of our original sample, leaving about $25\%$ of the original stellar pairs\footnote{Improved parallax inferences in future data releases will reduce the impact of the distance selection cut, yielding a larger effective pair catalog and thus potentially a faster improvement with integration time $\tau$, as compared to the estimates in Eqs.~\eqref{eq:snr_ss_mu},~\eqref{eq:snr_ss_alpha}, and~\eqref{eq:snr_ss_alpha_2}.}.

\subsubsection{Background motion subtraction and outliers removal} \label{sec:ss_back_sub}

\begin{figure*}   
    \includegraphics[width=1\textwidth]{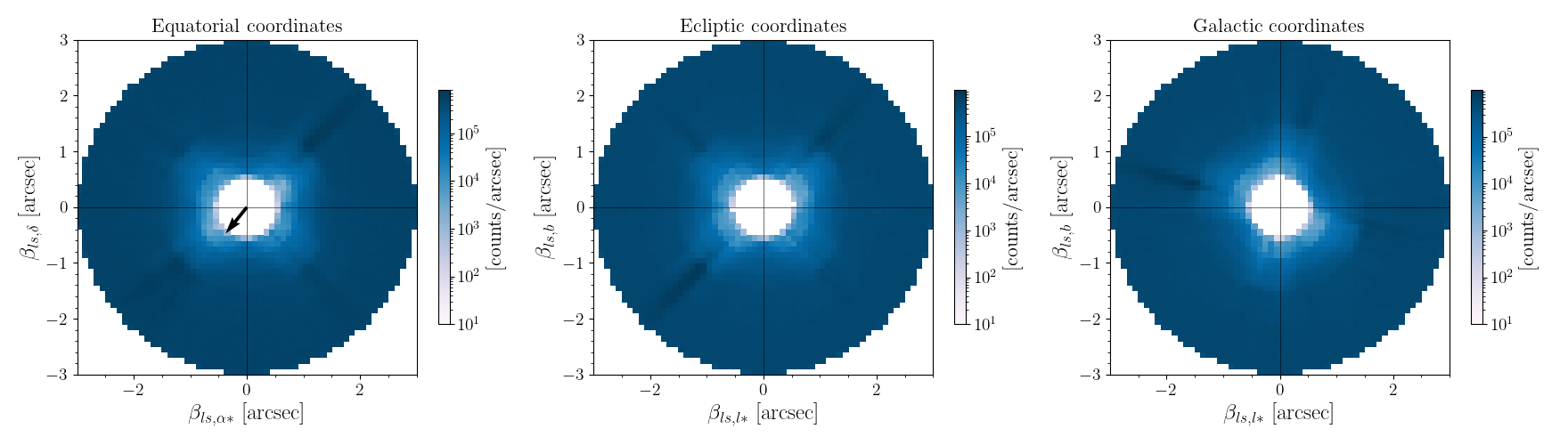}
    \includegraphics[width=0.685\textwidth]{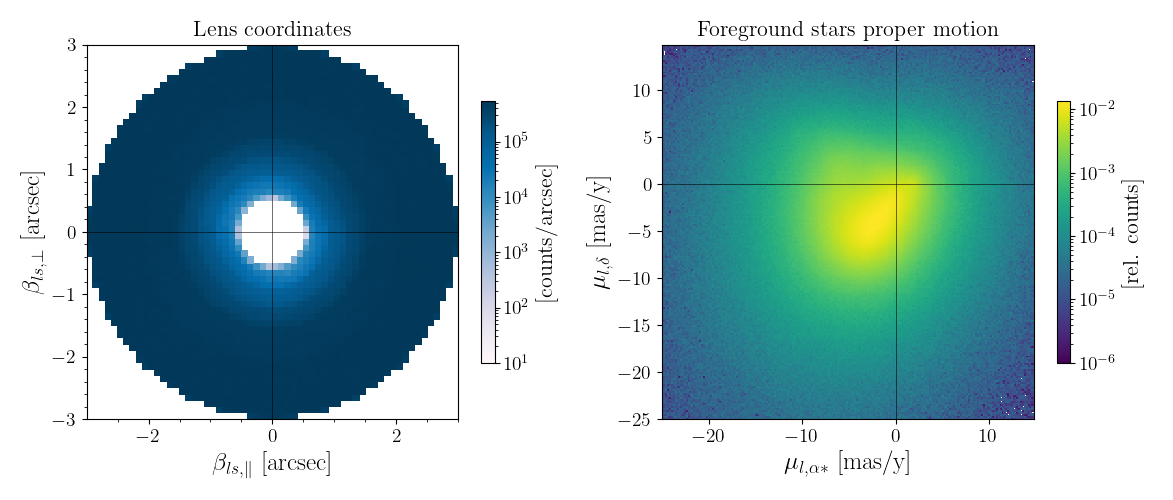}
    \caption{({\it Top row}) Number density of background stars around their lens counterpart in equatorial (left), ecliptic (center), and galactic (right) coordinates. For all stellar pairs, the lens is at the origin, and the black arrow in the left panel points in the direction of the average lens velocity over all pairs. The number of background stars decreases significantly below $\sim 1.5''$ and it is not uniformly distributed. ({\it Bottom left}) Number density of background stars around their lens counterpart as a function of the parallel and perpendicular components of the angular impact parameter with respect to the lens velocity direction where, for all stellar pairs, the lens is at the origin; in this plane, all the lens velocities are directed along the positive horizontal axis and the anisotropy is not visible, suggesting that it is not related to the lens velocity direction. ({\it Bottom right}) Distribution of the proper motions of the star-lenses. \nblink{star-star/make_plots.ipynb}}	
    \label{fig:aniso}
\end{figure*}

In order to avoid spurious signals contributing to the the proper motion template of Eq.~\eqref{eq:tau_mu_starstar}, we should make sure that $\mathcal{T}^{\star}_\mu$ vanishes in the absence of a real lensing signal. In the limit of background stars uniformly distributed around the lenses, we expect that $\langle \mathcal{T}^{\star}_\mu \rangle_{\rm{noise}} = 0$, owing to the parity symmetry of the dipole pattern of the lens-induced distortion $\Delta \vect{\mu}_{ls}$ of Eq.~\eqref{eq:mu_starstar}. This remains true even with a preferred lens velocity direction, $\langle \hat{\vect{\mu}}_{ls} \rangle \neq 0$, and a nonzero background star motion, $\langle \vect{\mu}^p_{s} \rangle \neq 0$, with averages taken over all pairs $p$. However, we find that the background stars are not distributed isotropically, but more background stars are observed in specific directions, particularly at small lens-source angular separation, as shown in the upper row of Fig.~\ref{fig:aniso} in equatorial, ecliptic, and galactic coordinates. The anisotropy does not seem to be (strongly) related to the foreground star's proper motion direction, as shown in the bottom left panel of Fig.~\ref{fig:aniso}, while the bottom right panel shows the distribution of the lenses' proper motion. We suspect that the most likely (and plausible) candidate for this anisotropy in number density can be attributed to \gaia's scanning law pattern. 

In order to mitigate the effect of the observed anisotropy, we subtract the mean motion of the background stars, which ensures that $\langle \mathcal{T}^{\star}_\mu \rangle_{\rm{noise}} = 0$ irrespective of the background stars' distribution. To do so, we bin the background stars in a 4-dimensional histogram based on their on-sky position, G magnitude, angular separation from their foreground counterpart, and inferred geometric distance. For the spatial pixelation, we use the nested HEALPix\footnote{http://healpix.sourceforge.net} scheme at level 4 \cite{Zonca2019, 2005ApJ...622..759G}, corresponding to a resolution of approximately $3.7\ \rm{deg}$. In G magnitude and angular separation, the bin sizes are chosen to be 1 and $0.3''$, respectively. Finally, due to the large uncertainties on the stars' distances, we use a probabilistic bin assignment in 4 logarithmically-spaced bins between 1 and 10 kpc (plus 2 additional bins for stars falling below and above this range); since \cite{2021AJ....161..147B} do not provide the full posterior distribution, we crudely model it as a Gaussian centered on the median, and with standard deviation $(\sigma_{l,\text{86th}}+ \sigma_{s,\text{14th}})/2$. To ensure that we can reliably estimate the statistics in each bin, we only retain stars with at least $80 \%$ of their probability support in bins with a threshold count of 30 stars. The removal of stars falling in ``sparse'' bins results in the reduction of the sample with spatial features due to the HEALPix pixelation, as shown in the right panel of Fig.~\ref{fig:doubles_map}. In each bin, we compute the proper motion mean $\bar{\vect{\mu}}$ and the effective proper motion dispersion covariance matrix $\Sigma_{\mu}$. Outlier stars with proper motion $\vect{\mu}$ too far from the mean $\bar{\vect{\mu}}$ -- those with $(\vect{\mu}-\bar{\vect{\mu}})\cdot \Sigma_{\mu}^{-1} \cdot (\vect{\mu}-\bar{\vect{\mu}})^T > 9$ -- are removed. Since the lensing signal is not expected to induce any such large deviation on individual stars, this cleaning ensures the removal of spurious fast-moving stars while retaining most of the signal. This procedure is iteratively repeated 10 times, with only about $0.01\%$ of outliers identified in the last iteration.  

\begin{figure*}   
    \includegraphics[width=0.49\textwidth]{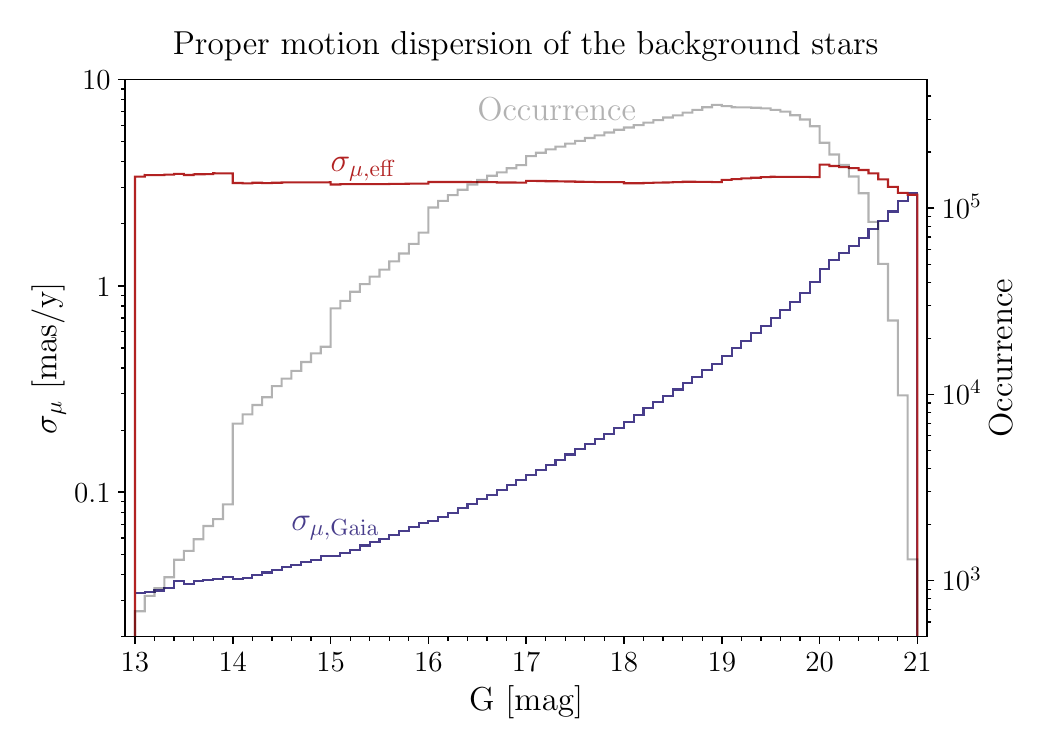}
    \includegraphics[width=0.49\textwidth]{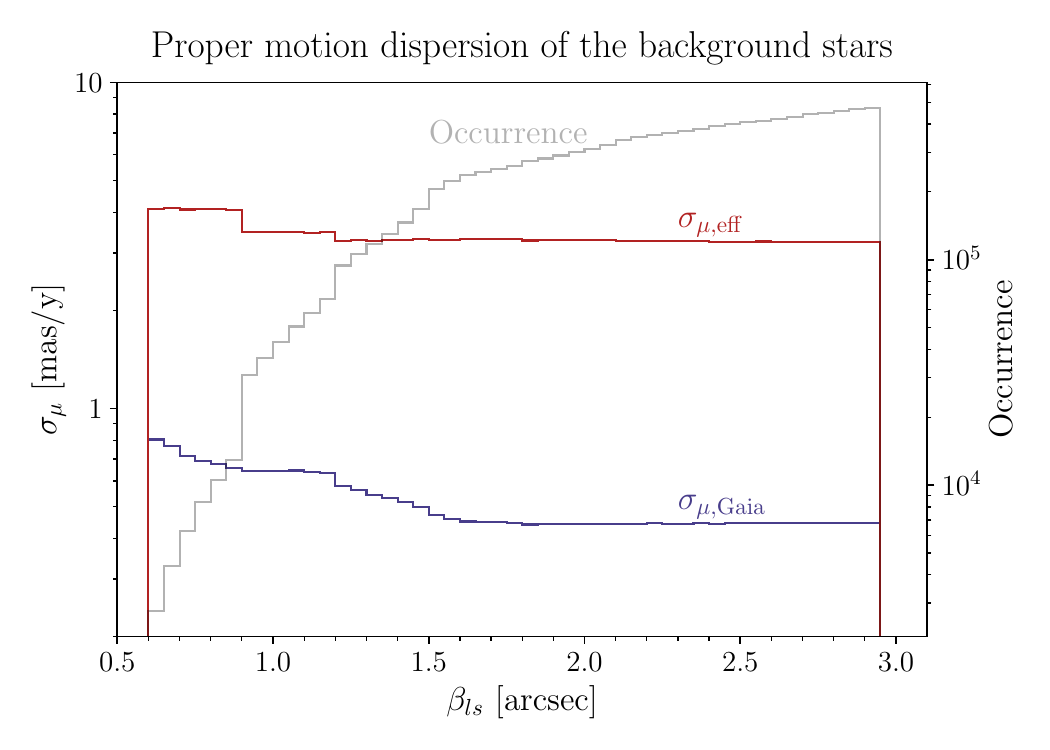}
    \caption{Proper motion dispersion of the background stars in the final sample of optical doubles stars in \gaia's EDR3, after quality and distance cuts, background motion subtraction and outlier removal (see sections~\ref{sec:quality_distance} and \ref{sec:ss_back_sub}). ({\it left}) Histograms of the number of stars per G magnitude bin (gray), average proper motion error reported by \gaia (blue), and average effective dispersion of the stars' proper motion in each bin (red). ({\it right}) Same as the left panel, but as a function of the background-foreground stellar angular separation. For both the effective and the \gaia error, the total proper motion variance is computed as $\sigma^2_{\mu} \equiv \sigma^2_{\mu_\alpha} + \sigma^2_{\mu_\delta} + 2 \rho_{\mu_\alpha \mu_\delta}\sigma_{\mu_\alpha}\sigma_{\mu_\delta} $, where $\rho_{\mu_\alpha \mu_\delta}$ is the correlation between the proper motion in the $\alpha$ and $\delta$ directions. \nblink{star-star/make_plots.ipynb}}	
    \label{fig:eff_error_starstar}
\end{figure*}

The final sample with the subtracted proper motion mean and the effective proper motion covariance can be used to compute the template of Eq.~\eqref{eq:tau_mu_starstar}. Altogether, about $77\%$ of the remaining stellar pairs survive the aforementioned cleaning procedure. We compare the estimated effective proper motion dispersion to the instrumental errors reported by \gaia as a function of the background stars' G magnitude and angular separation from their foreground counterpart in Fig.~\ref{fig:eff_error_starstar}. The observed dispersion of the stellar population is much larger than the error reported by \gaia and nearly independent of G magnitude, showing that for the majority of the background stars in our sample the instrumental precision is well below the intrinsic proper motion dispersion, as expected. The right panel of Fig.~\ref{fig:eff_error_starstar} shows that background stars which are closer to their foreground counterpart are more poorly measured and have larger dispersion, since they are observed in crowded regions; on the other hand, for angular separations $\gtrsim 1.5 ''$ the distributions are nearly flat. In the next section, we report results of the proper motion template analysis on the clean sample of optical double stars.

\subsection{Results and outlook}  \label{sec:star-star_res}

\subsubsection{Unweighted proper motion template}

The computation of the proper motion template from Eq.~\eqref{eq:tau_mu_starstar} on the cleaned sample of about 11.7 million stellar pairs leads to the results shown in Tab.~\ref{tab:star_template}. All stellar lenses are equally weighted with unit Solar mass. In addition to the signal channel, which computes the overlap of the background stars' proper motion with the lens-induced dipole profile, we consider three control channels: a dipole profile rotated by $90^{\circ}$, a monopole profile $\Delta \vect{\mu}_{ls} \propto \vect{\hat{\beta}}_{ls}$, and a quadrupole profile $\Delta \vect{\mu}_{ls} \propto  2 \muhat_{ls} \left( \betahat_{ls} \cdot \muhat_{ls} \right) + \betahat_{ls} \left( 1- 4(\betahat_{ls} \cdot \muhat_{ls} )^2 \right)$, all with the same radial scaling as the signal dipole channel. The expected SNR is similar in all channels and below unity, as expected, about a factor of 4 smaller than our naive estimate of Eq.~\eqref{eq:snr_ss_mu}. 

\begin{figure*}   
    \includegraphics[width=0.98\textwidth]{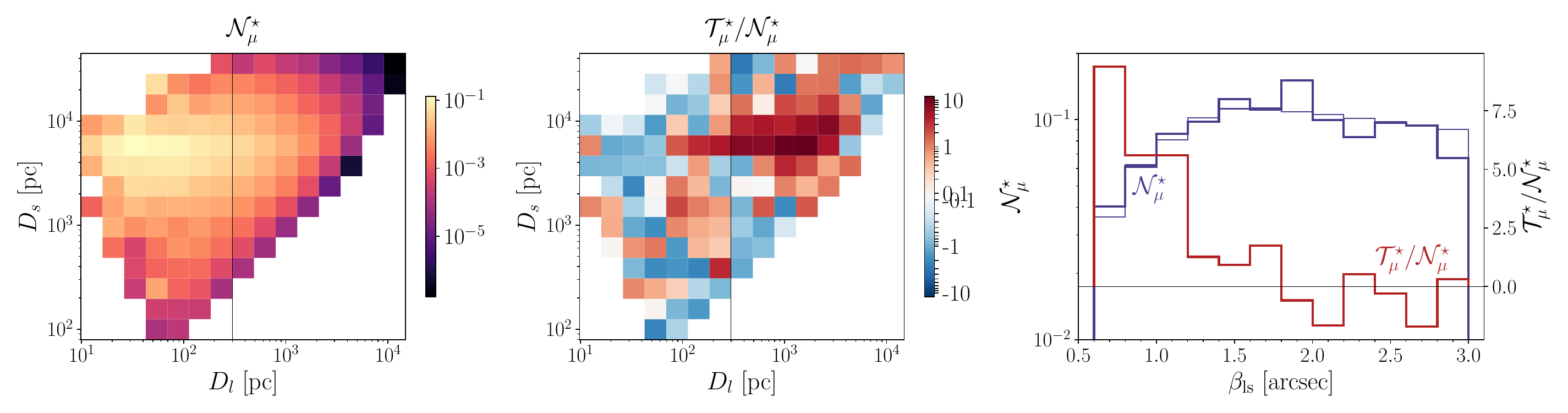}
    \caption{({\it left}) Expected SNR, $\mathcal{N}^{\star}_{\mu}$, for the star-star proper motion template on the clean sample of 11.7 million pairs as a function of lens and source distance. ({\it center}) Same as the left panel but for the observed $\mathcal{T}^{\star}_{\mu}/\mathcal{N}^{\star}_{\mu}$. The region with $D_l < 300\ \rm{pc}$ contains most of the sensitivity and shows the smallest fluctuations in $\mathcal{T}^{\star}_{\mu}/\mathcal{N}^{\star}_{\mu}$; for $D_l > 300\ \rm{pc}$ we observe high positive fluctuations around $D_s \simeq 7\ \rm{kpc}$ which are most likely due to a systematic error that has not been removed by the cleaning procedure. ({\it right}) Expected SNR, $\mathcal{N}^{\star}_{\mu}$ (thick blue, left axis), and observed SNR, $\mathcal{T}^{\star}_{\mu}/\mathcal{N}^{\star}_{\mu}$ (red, right axis), as function of the lens-source angular separation $\beta_{\rm{ls}}$. The thin blue line denotes the expected scaling of the SNR as $\sqrt{N_{\rm{pairs}}}/(\beta_{ls}\sigma_{\mu, \rm{eff}})$ and is arbitrarily normalized to approximately match the bins with the largest SNR. Here $N_{\rm{pairs}}$ and $\sigma_{\mu, \rm{eff}}$ denote the total number of pairs and the effective dispersion averaged over all the background stars in each bin. \nblink{star-star/make_plots.ipynb}}	
    \label{fig:starstar_res}
\end{figure*}

The dipole signal channel shows an excess of $3.75 \sigma$. However, at this point, we caution against interpreting such measurement as a positive detection -- and attribute it to a potentially-resolvable systematic -- for the following reasons. Taken at face value, the measured value of $\mathcal{T}^{\star}_{\mu}$ corresponds to an average lens mass of $11\ M_{\odot}$, which is incredulously high given \gaia's estimated astrophysical parameters, as discussed below. Moreover, we further test the signal hypothesis by evaluating the template on subsamples of the entire catalog of optical doubles (see Fig.~\ref{fig:starstar_res}), in which the pairs are binned according to the foreground and background distances (left and middle panel) and their angular impact parameter (right panel). The expected scaling of $\mathcal{N^\star_\mu}$ with $D_l$ and $\beta_{ls}$ is borne out in the data. The peaks in $\mathcal{T^\star_\mu}/\mathcal{N^\star_\mu}$ occur in bins with the smallest SNR, contrary to the expectation from a real signal. We conclude that the large $\mathcal{T}^{\star}_{\mu}$ value reported in Tab.~\ref{tab:star_template} is most likely due to a systematic effect that has not been properly removed by our cleaning procedure and that is more pronounced in bins with poor statistics. The middle panel of Fig.~\ref{fig:starstar_res} reveals how the anomalously-large $\mathcal{T^\star_\mu}$ value is mostly due to optical doubles where the lens is further than about $300\ \rm{pc}$, where the SNR is below $0.07$. This justifies removing pairs with $D_l > 300\ \rm{pc}$ from our sample, reducing the size to about 0.814 million stellar pairs that retain most of the sensitivity. As shown in Tab.~\ref{tab:star_template_close}, the value of $\mathcal{N^\star_\mu}$ is almost unchanged and the measured $\mathcal{T^\star_\mu}$ does not show a significant fluctuation, as expected for the (statistical-)background-only hypothesis. The removal of distant lenses is also justified by the sensitivity estimates discussed above (see Eq.~\eqref{eq:snr_ss_mu}), which showed how the SNR is maximized by lenses at $D_{l} \simeq v_{ls}\tau/\beta_{\rm{min}} \simeq 270\ \rm{pc}$, for $v_{ls} \simeq 10^{-3}c$, $\tau = 3\ \rm{y}$, and $\beta_{\rm{min}} = 0.7''$.

\begin{table} 
\renewcommand{\arraystretch}{1.2} 
\begin{center}
\begin{tabular}{c c c c c} 
    Matched-filter & $\mathcal{T}^{\star}_{\mu}$ & $\mathcal{N}^{\star}_{\mu}$ & $\mathcal{T}^{\star}_{\mu}/\mathcal{N}^{\star}_{\mu}$ & $\langle m_l \rangle\ [M_\odot]$ \vspace{0.1cm} \\
    \hline \hline 
     Dipole & 1.25 & 0.34 & 3.75 & 11 \\
     \hline
     Dipole ($90^\circ$) & 0.31 & 0.36 & 0.86 & - \\
     \hline 
     Monopole & 0.74 & 0.34 & 2.16 & -\\
     \hline
     Quadrupole & -0.05 & 0.34 & -0.14 & -\\
\end{tabular}
\caption{\label{tab:star_template} Proper motion template evaluated on the sample of 11.7 million cleaned \gaia DR3 optical doubles. The first row corresponds to the signal channel, where the observed background stars' proper motion are matched filtered with the lensing-induced dipole profile, while the other rows correspond to control channels, where no signal is expected. In each case, the template, its normalization (expected SNR) and their ratio (observed SNR) are reported. For the signal channel, the measured average lens mass is also indicated. \nblink{star-star/compute_template.ipynb}}
\end{center}
\end{table}

\begin{table} 
\renewcommand{\arraystretch}{1.2} 
\begin{center}
\begin{tabular}{c c c c c} 
    Matched-filter & $\mathcal{T}^{\star}_{\mu}$ & $\mathcal{N}^{\star}_{\mu}$ & $\mathcal{T}^{\star}_{\mu}/\mathcal{N}^{\star}_{\mu}$ & $\langle m_l \rangle\ [M_\odot]$ \vspace{0.1cm} \\
    \hline \hline 
     Dipole & 0.2 & 0.33 & 0.61 & 1.9 \\
     \hline
     Dipole ($90^\circ$) & 0.26 & 0.36 & 0.73 & - \\
     \hline 
     Monopole & 0.39 & 0.33 & 1.2 & -\\
     \hline
     Quadrupole & -0.06 & 0.34 & -0.17 & -\\
\end{tabular}
\caption{\label{tab:star_template_close} Same as Tab.~\ref{tab:star_template} but for a sample of 814 thousand cleaned \gaia DR3 optical doubles with lenses closer than $300\ \mathrm{pc}$ to the observer. \nblink{star-star/compute_template.ipynb}}
\end{center}
\end{table}

Recently \cite{mcgill2020predictions} pointed out that a significant fraction of microlensing events predicted in {\it Gaia} DR2 were spurious events, where the background source candidate turned out to be either a duplicate detection or a binary companion of the lens. The issue was found to mostly affect bright sources with no 5D astrometric solution. We believe it is unlikely that a similar spurious contamination affects our sample of lens-source pairs for the following reasons. First, by construction, we only select background sources with proper motion measurement and we additionally require high quality of the astrometric solution, together with significant difference in the measured parallaxes and distances in each pair (see Sec.~\ref{sec:quality_distance}). These requirements ensure that the majority of the pairs are genuine sources that are widely separated along the line of sight. Nevertheless, we performed sanity checks on the distribution of source-lens difference in magnitude and color, following the procedure adopted by \cite{mcgill2020predictions}, and found no anomalous clustering of photometrically identical pairs, particularly for the not-super-faint sources that were identified as problematic in the microlensing events. Based on these results, we are confident that the sample of about 814 thousand pairs that drives the sensitivity to the astrometric lensing signal is reliable.

\subsubsection{Alternative templates} \label{sec:star_astro_param}

\gaia DR3 provides astrophysical parameters from BP/RP and RVS spectra for nearly 0.5 million sources \cite{2023A&A...674A..26C}, \cite{2023A&A...674A..28F}, \cite{2023A&A...674A..31D}, including an inferred mass for  128 million stars. In principle, the mass estimate for the foreground stars in our sample -- or a proxy for the mass, such as the luminosity or the effective temperature -- can be used as an additional weight for the lensing correction in the template of Eq.~\eqref{eq:tau_mu_starstar}.
From our sample of 11.7 million clean pairs, around 2 million lenses have an estimated mass. The SNR of this subsample (weighting all the lenses equally with a mass of $M_{\odot}$) is reduced by a factor of 0.13 with respect to the full sample, which is 3 times worse than the naive scaling with $\sqrt{N_{\rm{pairs}}}$. This is likely due to the fact that inferring astrophysical parameters is harder in crowded regions, where the lensing signal is maximized. If we include the inferred mass as weights, we get $\mathcal{N}^{\star}_{\mu} = 0.03$ instead of $\mathcal{N}^{\star}_{\mu} = 0.046$ with unit weights, a reduction that is expected given the lenses' median mass of $0.89\ M_{\odot}$. 

Given the reduced sensitivity, we refrain from using astrophysical parameters at this point. This is further justified by the well known issue of \gaia's photometry in crowded regions: the mean source photometry is more likely to be blended with other sources in all or some of the observation epochs, possibly having a significant effect on the estimated photometry \cite{2021A&A...649A...3R}. In future data releases, the effect of crowding on the spectra will be mitigated, opening up the possibility of reliably using astrophysical parameters for stars with an accidentally-close pair that can induce a lensing signal.

The time-varying lensing effects induced by the foreground stars could also manifest as an induced parallax in the background stars that can be captured by an appropriate template observable, as discussed in Sec.~\ref{sec:dm_templ} and worked out in detail in appendix~\ref{app2:parallax}. We did not include it in our current analysis due to the challenges encountered when modeling and subtracting the background, which is crucial to remove large systematic errors. Differently from the stellar proper motions, the distribution of observed parallaxes is very different from a Gaussian distribution and highly skewed. This makes it difficult to subtract the mean background parallax and model the effective dispersion, which are needed in order to evaluate the significance of the observed value of $\mathcal{T^{\star}_\varpi}/\mathcal{N^{\star}_\varpi}$. Since the parallax template is expected to be less sensitive than the proper motion template (or at most comparable) and the signal is currently below the detection threshold, we leave further studies of the parallax background to future work.

\section{Conclusions}
\label{sec:conclusion}
The precision astrometry provided by {\it Gaia} has opened several windows for understanding the dark matter that makes up the Milky Way halo. Aside from singular astrometric microlensing events, the precision of {\it Gaia} allows one to consider aggregate statistical observables as well. 

One class of such observables is the {\em matched-filter} or template observable to look for correlated motions or accelerations of background stars induced by foreground lenses \citepalias{2018JCAP...07..041V}. Here, while individual stars may not provide a signal above noise levels, fitting a combination of luminous sources can yield a signal, or provide constraints. The first search using these techniques was performed by \citetalias{2020PhRvL.125k1101M}, finding no signal as expected, even assuming all of the dark matter was in the form of such objects. 

In this work, we have extended those analyses using DR3, placing limits on dense subhalos with scale radii of $1\ \textrm{pc}$ and $10^{-3}\ \textrm{pc}$, with constraints on order-unity fractions of the dark matter in such objects down to halo masses of $2 \times 10^{7}\ M_\odot$ and  $10^{6}\ M_\odot$, respectively. Such limits can yet improve by an order of magnitude or more in future {\it Gaia} data releases before intrinsic stellar proper motion dispersion limits further improvement.

We have also applied the velocity template to star-star pairs. While astrometric lensing by individual stars is too weak to generate a detectable signal on a companion star, it is possible that by aggregating the signals from many stars, we may see this effect. We have created a catalog of 61 million optical doubles, from which we have extracted a cleaned subsample of 11.7 million, with 814 thousand pairs in our expected signal region. As expected, we find no excess in our main signal region, while we find a small excess correlation (inconsistent with a real signal) in the full cleaned sample. The origin of the latter spurious fluctuation remains unknown, but intrinsic correlation of stellar motions may play some role. While we do not expect velocity templates to provide a measurement of masses given the intrinsic dispersion of stellar motions, acceleration templates should be able to detect a signal and perhaps serve as a tool to measure properties of stars.

Going beyond velocity templates, we have further extended the matched-filter approach to accelerations and parallax. While accelerations were previously noted as a possible template in \citetalias{2018JCAP...07..041V}, no sensitivity analyses were performed. We have forecast the sensitivity of acceleration templates and found they can be a powerful probe in the range of $1 \ M_\odot - 10^7\ M_\odot$. While {\it Gaia} does not provide acceleration data at this time, one can extract accelerations analytically using a combination of DR2 and DR3 data. We have forecast the sensitivity for this approach and find that even this could provide interesting limits on dark halos in the mass range $10^3 \ M_\odot - 10^7\ M_\odot$.

We have also studied for the first time the time-dependent lensing effect arising from the parallax motion of the intervening lens. The parallax motion of the lens can induce an anomalous parallax motion on a lensed object, yielding motions that have annual periodicity, but can, in general, be quite different from true parallax. As these motions are not currently included in \gaia's astrometric solution, we develop a template that extracts the component of parallax that would be found when fitting to conventional parallax. The test statistic from this template does not improve as rapidly with time as velocity or acceleration (both of which naturally improve from the extended time baseline, alone) and thus is not expected to be a significant contributor to limits or discovery in the future. However, such an effect could be useful in looking for ``outlier'' velocities and accelerations as suggested in \citetalias{2018JCAP...07..041V}, distinguishing them from non-lens sources of motion.

In summary, it is clear that template observables of time-domain lensing are approaching an exciting era, where measured velocities and accelerations will probe regions of parameter space for dark halos that are otherwise unexplored. In addition, there is the possibility to measure this effect in large sets of star-star pairs in the near future. Anomalous parallax may prove a useful tool in future analyses as {\it Gaia} continues to provide subtle insights into the nature of the dark matter in the Milky Way.
 
\section*{Acknowledgements}
KVT is supported by the National Science Foundation under Grant PHY-2210551. NW is supported by NSF under Award No. PHY-2210498, by the BSF under Grant No. 2018140. NW and AT are supported by the Simons Foundation. 
Some of the results in this paper have been derived using the healpy and HEALPix packages.
This work has made use of data from the European Space Agency (ESA) mission \gaia (\url{https://www.cosmos.esa.int/gaia}), processed by the \gaia Data Processing and Analysis Consortium (DPAC, \url{https://www.cosmos.esa.int/web/gaia/dpac/consortium}). Funding for the DPAC has been provided by national institutions, in particular the institutions participating in the \gaia Multilateral Agreement.
This work was supported in part through the NYU IT High Performance Computing resources, services, and staff expertise. The Center for Computational Astrophysics at the Flatiron Institute is supported by the Simons Foundation. Research at Perimeter Institute is supported in part by the Government of Canada through the Department of Innovation, Science and Economic Development Canada and by the Province of Ontario through the Ministry of Economic Development, Job Creation and Trade. 

%%%%%%%%%%%%%%%%%%%%%%%%%%%%%%%%%%%%%%%%%%%%%%%%%%
\section*{Data Availability}

The data used in this article are publicly available on the \href{https://gea.esac.esa.int/archive/}{\gaia Archive}. Furthermore, the processed data produced in the analysis are available at \href{https://users.flatironinstitute.org/~kvantilburg/lens-velocity/}{this link}. The codes to reproduce the proper motion template analysis on the MCs (Sec.~\ref{sec:veltempl_update}) and the star-star lensing template analysis (Sec.~\ref{sec:star-star_data}) are available on GitHub \githubmaster.

	
	
	\bibliographystyle{mnras}
	\bibliography{ref}

	%\begin{thebibliography}{99}
	%\end{thebibliography}
	
	%%%%%%%%%%%%%%%%%%%%%%%%%%%%%%%%%%%%%%%%%%%%%%%%%%
	
	%%%%%%%%%%%%%%%%% APPENDICES %%%%%%%%%%%%%%%%%%%%%
	
	\appendix

\section{Test statistic for multiple lenses} \label{app1:multi-lens}

The global test statistic $\mathcal{R}$ used in the analysis of section~\ref{sec:veltempl_update} takes into account only the dominant contribution to the signal, coming from the lensing distortion induced by the lens that is closest to the observer -- see Eq.~\eqref{eq:snr}. Typically, however, there will be many more lenses in front of the stellar target that are further away and therefore produce a progressively weaker -- but possibly still detectable -- signal. In this appendix, we consider a generalization of the test statistic that accounts for such subleading contributions in the limit that the lenses in the FOV are sparse, i.e.~non overlapping. In particular, we would like to compare the sensitivity of the single-lens observable used so far with a more general multi-lens observable. The results presented below show how the adopted test statistic already has the best sensitivity compared to the simplest multi-lens generalization, assuming our conservative treatment from the main text. The main reason is that we do not attempt to model the noise faithfully over the entire region, and (in our signal simulations) only look at contributions to the test statistic around actual lenses. Further optimizations towards better noise models and multi-lens statistics are left to future work. 

We start by recalling the definition of the global test statistic $\mathcal{R}$ obtained from maximizing the likelihood ratio of the signal hypothesis over the background-only hypothesis assuming Gaussian uncorrelated proper motion noise, and retaining only the strongest signal from the lens that is closest to the observer:\footnote{See the Supplemental Material of \citetalias{2020PhRvL.125k1101M} for its derivation.}
%\begin{widetext}
\begin{align} \label{eq:test_statistic_R}
    & \mathcal{R}(M_l, r_l, f_l | \lbrace \vect{\mu}_i \rbrace) =  \nonumber \\
    & \sup_{\vect{\theta}_t, \beta_t} \Bigg [ \ln \frac{\rho_l(\vect{\theta}_t, r_l/\beta_t)}{\beta_t^4} + \frac{C^2 \sigma_v^2 \mathcal{N}^2\Big(\frac{\mathcal{T}^2}{\mathcal{N}^2}  -  \frac{v_0^2}{\sigma_v^2}  \Big)+ 2 C \vect{\mathcal{T}}	\cdot\vect{v}_0}{2(1+C^2 \sigma_v^2 \mathcal{N}^2)} \Bigg ],
\end{align}
%\end{widetext}
where $C = 4 G M_l/ (c^2 r_l^2)$, $v_0$ is the magnitude of the observer velocity projected along the line of sight, and $\sigma_v$ is the variance of the lens velocity Gaussian distribution. In the expression above $\vect{\mathcal{T}} \equiv \left\lbrace \mathcal{T}_\mu(\beta_t,  \vect{\theta}_t, \hat{\vect{\alpha}}), \mathcal{T}_\mu(\beta_t,  \vect{\theta}_t, \hat{\vect{\delta}}) \right\rbrace$ and $\mathcal{T}^2 \equiv |\vect{\mathcal{T}}|^2$. The SNR of the above test statistic can be estimated analytically within some simplifying assumptions. To this end, we neglect marginalization over $\beta_t$\footnote{The value of $\beta_t$ used for the analysis of a given parameter space point $(M_l, r_l, f_l)$ is fixed to be approximately $\beta_{t, \rm{opt}} \simeq r_l/D_3$, where $D_3$ is the distance from the observer within which there are at least 3 lenses, implicitly defined as $\Delta\Omega \int_0^{D_3} \dd D \, D^2 \rho_l(D)/M_l = 3 $.}, %$D_3 \simeq \left[ 9M_l/(\Delta\Omega \rho_l) \right]^{1/3}$, for a const.
and work in the limit of a uniform field of stars $\Sigma_0/\sigma_{\mu} = \rm{const.}$ A crucial difference between the global observable $\mathcal{R}$ and the local template $\mathcal{T}$ is the marginalization over the unknown lens location $\vect{\theta}_t$. In other words, $\mathcal{R}$ approximately gives the largest of the many $\mathcal{T}$ values resulting from scanning over the stellar target's solid angle $\Delta \Omega$ in a 2D grid with a lattice constant approximately equal to $\beta_t$. The number of independent ``trials'' is thus
\begin{align}
    N_{\rm{trials}} \simeq \frac{\Delta \Omega}{\beta_t^2} \simeq 10^5 \frac{\Delta \Omega}{(\ang{10})^2}\left(\frac{\ang{;;100}}{\beta_t}\right)^2,
\end{align} 
where the above numerical values are typical for the LMC analysis. In the presence of a visible lensing signal, the largest $\mathcal{T}$ is obtained at the location of the lens. In the absence of a signal, the global test statistic is instead given by the largest statistical fluctuation in $N_{\rm{trials}}$ samples of a random variable. Therefore, while $\langle \mathcal{T} \rangle_{\rm{noise}} = 0$, the typical $\mathcal{R}$ does not vanish for noise only. Because of this look-elsewhere effect, the single-lens observable will turn out to be more sensitive than the multi-lens one, as we show below.

It is straightforward to generalize the $\mathcal{R}$ test statistic to include the contribution from $n_l$ lenses in front of the stellar target in the limit that they do not overlap, i.e.~when $\Delta \Omega \gg n_l \beta_t^2$,
\begin{align} \label{eq:multilens_repl}
    \mathcal{R}_{\rm{multi-lens}} = \ln \frac{r_l^{3n_l}}{n_l! M_l^{n_l}} -\langle N_l\rangle + \sum_{t=1}^{n_l} \mathcal{R}_t,
    %\sup_{\vect{\theta}_t, \beta_t}[] \rightarrow \frac{\langle N_l\rangle^{n_l}}{n_l!} e^{-\langle N_l\rangle} \sum_{t=1}^{n_l}
\end{align}
where $\langle N_l\rangle = \int \dd^3r\ \rho_l(r)/M_l$ is the average number of lenses in the FOV and $\mathcal{R}_t$ denotes the quantity in Eq.~\eqref{eq:test_statistic_R}. For $t>1$, the maximum over $\vect{\theta}_t$ can be obtained excluding a region of a few $\beta_t$s from the maxima found for lenses from $1$ to $t-1$. For a population of lenses distributed with an approximately constant galactic density profile, the distance to the $t$-th lens scales as $D_t = t^{1/3} D_1$, where $D_1$ is the distance to the closest lens given by Eq.~\eqref{eq:Dl_min}. We neglect marginalization over the templates' angular size and assume that they can be fixed by following the lens distance scaling, $\beta_t = \beta_1/t^{1/3}$. Accounting for multiple lenses introduces the additional parameter $n_l$ in the likelihood \eqref{eq:multilens_repl}. For simplicity, here we fix $n_l = \langle N_l\rangle$. We can now estimate the SNR of the test statistic working in two limiting cases: $v_0 \gg \sigma_v$, i.e.~when all the lenses have the same velocity $\vect{v}_0$, and $\sigma_v \gg v_0$, i.e.~when the lenses' velocities are uncorrelated. 

{\it Fixed lens velocity} ($v_0 \gg \sigma_v$): in this case Eq.~\eqref{eq:test_statistic_R} reduces to
\begin{align} \label{eq:test_statistic_R_case1}
    \mathcal{R}^{v_0 \gg \sigma_v}_t & \simeq C v_0 \sup_{\vect{\theta}_t} \Bigg[ \mathcal{T}_t(\beta_t,  \vect{\theta}_t, \hat{\vect{v}}_0) \Bigg ] + \rm{const.},
\end{align}
where only the term that depends on the data has been kept explicitly. $\mathcal{R}_t$ follows an extreme value distribution of the Gaussian variable $\mathcal{T}_t$, which is approximately a Gumbel distribution for large $N_{\rm{trials}}$. Therefore, the average expectation value over noise is
\begin{align}
    \langle \mathcal{R}^{v_0 \gg \sigma_v}_{\rm{multi-lens}} \rangle_{\rm{noise}} & \simeq C v_0 \sum_{t} f(N_{\rm{trials}}-t) \mathcal{N}_t + \rm{const.},
\end{align}
where 
\begin{align} \label{eq:ffunc}
    f(N) & = \sqrt{2} \left[ (1-\gamma)\ \mathrm{erf}^{-1}\left(1-\frac{2}{N}\right) + \gamma\ \mathrm{erf}^{-1}\left(1-\frac{2}{e N}\right) \right] \nonumber \\
         & \simeq (1-\gamma)\sqrt{\ln{\frac{\widetilde{N}}{\ln{\widetilde{N}}}}} + \gamma \sqrt{2+\ln{\frac{\widetilde{N}}{2+\ln{\widetilde{N}}}}},  
\end{align}
$\gamma$ is the Euler constant, $\rm{erf}^{-1}$ is the inverse error function and $\widetilde{N} \equiv N^2/(2\pi)$. On the other hand, in the presence of a lensing signal \begin{align}
    \langle \mathcal{R}^{v_0 \gg \sigma_v}_{\rm{multi-lens}} \rangle_{\rm{sign}} & \simeq C^2 v^2_0 \sum_{t} \mathcal{N}^2_t + \rm{const.}
\end{align}
We can then estimate the SNR using also the variance $\langle (\mathcal{R}^{v_0 \gg \sigma_v}_{\rm{multi-lens}})^2 \rangle = C^2 v_0^2 \sum_t \mathcal{N}^2_t$. The requirement of $\SNR > 1$ is equivalent to
\begin{align} \label{eq:snrreq_multi_1}
    C v_0 \mathcal{N} & > \frac{1}{\sqrt{\sum_t t^{-2/3}}} + \frac{\sum_{t} f(N_{\rm{trials}}-t) t^{-1/3}}{\sum_t t^{-2/3}},
\end{align}
where $\mathcal{N}$ corresponds to the closest lens and the lens distance scaling has been used to replace $\mathcal{N}_t = \mathcal{N}t^{-1/3}$. The first term on the right-hand side of Eq.~\eqref{eq:snrreq_multi_1} is the improvement factor in sensitivity that one would expect naively when adding up the signal from multiple lenses. However, due to the look-elsewhere effect, there is an additional contribution from the second term which increases with the number of lenses due to the weak scaling of the function $f$. Overall, the condition above requires a stronger signal than the detection of an individual lens
\begin{align} \label{eq:snr_req_1}
    C v_0 \mathcal{N} & > 1 +f(N_{\rm{trials}}) \simeq 1 + 4.4\ \frac{f(N_{\rm{trials}})}{f(10^5)}
\end{align}

{\it Uncorrelated lenses' velocities} ($\sigma_v \gg v_0$): in this case the global test statistic reduces to
\begin{align} \label{eq:test_statistic_R_case2}
    \mathcal{R}^{\sigma_v \gg v_0} & \simeq 
     \frac{1}{2}\sup_{\vect{\theta}_t} \Bigg [\frac{\mathcal{T}^2}{\mathcal{N}^2} \Bigg ] + \rm{const.},
\end{align}
assuming $1 + C^2 \sigma_v^2 \mathcal{N}^2 \simeq C^2 \sigma_v^2 \mathcal{N}^2$ in the denominator of Eq.~\eqref{eq:test_statistic_R}. For noise, $\mathcal{T}^2/\mathcal{N}^2$ follows a $\chi^2$-distribution with 2 degrees of freedom, so the average extreme value from $N_{\rm{trials}}$ is
\begin{align}
    \langle \mathcal{R}_{\rm{multi-lens}}^{\sigma_v \gg v_0} \rangle_{\rm{noise}} & \simeq \sum_t \left[ \gamma + \ln{ \left(N_{\rm{trials}}-t \right)} \right]  + \rm{const.} 
\end{align}
On the other hand, when there is a lensing signal, we will obtain the templates evaluated at the location of the lenses 
\begin{align}
    \langle \mathcal{R}_{\rm{multi-lens}}^{\sigma_v \gg v_0} \rangle_{\rm{sign}} & \simeq \sum_t \left[ \frac{C^2 \langle v_{l \perp} \rangle^2 \mathcal{N}^2_t}{2}  + 1  \right] + \rm{const.},
\end{align}
where we can use $\langle v_{l \perp} \rangle = \sqrt{\pi/2}\sigma_v$. In this case the variance is simply one and the condition $\SNR > 1$ is met when 
\begin{align} \label{eq:eq:snrreq_multi_2}
    C \langle v_{l \perp} \rangle \mathcal{N}  > \sqrt{2\left[\frac{1}{\sum_t t^{-2/3}} - \frac{(1 -\gamma )\langle N_l \rangle }{\sum_t t^{-2/3}} + \frac{\sum_t \ln{ \left(N_{\rm{trials}}-t \right) } }{\sum_t t^{-2/3}}  \right]},
\end{align}
where again $\mathcal{N}$ corresponds to the closest lens and we used $\mathcal{N}_t = \mathcal{N}t^{-1/3}$. The first term on the right-hand side of Eq.~\eqref{eq:snrreq_multi_1} gives the improvement expected without the look-elsewhere effect. However, due to the weak scaling of the last term, the right-hand side of the equation above increases with the number of lenses. Therefore, it is more convenient to include only the closest lens which can be detected if 
\begin{align} \label{eq:eq:snr_req_2}
    C \langle v_{l \perp} \rangle \mathcal{N} > \sqrt{2 \left(\gamma + \ln{N_{\rm{trials}}} \right)} \simeq 4.9 \frac{\sqrt{\gamma + \ln{N_{\rm{trials}}} }}{\sqrt{\gamma + \ln{10^5} }}.
\end{align}

The results above show that adding up the contribution from multiple, non-overlapping lenses in front of the stellar target does not improve the sensitivity of our analysis. This holds true in both of the limiting cases considered, and we therefore expect it to be a general result for any value of the parameters of the lens velocity distribution, $v_0$ and $\sigma_v$. The worst sensitivity of the test statistic in Eq.~\eqref{eq:multilens_repl} is due to the large number of tentative lens locations that need to be scanned over, which decreases weakly with each additional lens since $N_{\rm{trial}} \gg \langle N_l \rangle$. The results presented here are no longer valid when the number of lenses becomes so large that they start overlapping and $N_{\rm{trial}} \sim \langle N_l \rangle$. In this latter regime, lensing effects are more easily distinguishable using observables which are different from the templates considered in this work, and leverage global correlations in the proper motion or acceleration fields, as described in \cite{Mishra-Sharma:2020ynk}.

\section{Anomalous parallax}
\label{app2:parallax}
	
In addition to the astrometric lensing corrections described in section~\ref{sec:dm_templ}, which are due to the linear motion of the lens, the parallax displacement of the lens with respect to the observer induces a parallax-like lensing effect that can also be captured by a template observable applied to a field of background sources. Here we derive the expression of this anomalous parallax and compare its sensitivity to the proper motion template described in the main text.

We neglect the intrinsic parallax of the sources, and consider only the change in relative position due to the lens parallax $\varpi_l$. For simplicity, we model the lens parallax motion assuming that the observer moves in a circular orbit with angular velocity $\omega \simeq 2\pi/\text{y}$. The time-varying contribution to the source-lens impact parameter is then
\begin{align} \label{eq:parallax_bil}
    \vect{\beta}^\varpi_{li}(t) \simeq  \varpi_l \left\lbrace \cos \omega t, \sin \delta_\text{ecl} \sin \omega t \right\rbrace \equiv \varpi_l \hat{\vect{\varpi}}_l ,
\end{align}
where $\delta_\text{ecl}$ is the ecliptic latitude of the lens.\footnote{ Notice that with this definition $\hat{\vect{\varpi}}_l$ is a unit vector only at the ecliptic poles $\delta_\text{ecl} = \pm\pi/2$.} Expanding the lensing angular deflection of Eq.~\eqref{eq:ang_defl} in the template approximation, $\varpi_l \ll \beta^0_{li}$, the leading-order time-dependent correction to the background source position is
\begin{align}\label{eq:anomalous_par}
    \Delta\vect{ \theta}_i(t) = - \frac{4 G M_l \varpi_l }{r_l \beta_l} & \left\lbrace \frac{\widetilde{M}_l(\beta)}{\beta^2 / \beta_l^2} \left[\betahat^\varpi_{li}(t) - 2(\betahat \cdot \betahat^\varpi_{li}(t))\betahat \right] +  \right. \nonumber \\
    & \phantom{+} \left. \frac{\partial_{\beta} \widetilde{M}_l(\beta)}{\beta/ \beta_l^2} (\betahat \cdot \betahat^\varpi_{li}(t))\betahat \right\rbrace.
\end{align}
For observational times longer than a year, this periodic motion is \emph{partially} captured by the astrometric fit as a contribution to the source's parallax $\Delta \varpi_{li}$. Since source and lens are at small angular separations, the source parallax motion will be approximately modelled as $\Delta \varpi_{li} \hat{\vect{\varpi}}_l$ and the value of $\Delta \varpi_{li}$ that is inferred from the fit can be found by minimizing $|\Delta\vect{ \theta}_{li}(t) - \Delta \varpi_{li} \hat{\vect{\varpi}}_l|^2$ and averaging over one period. The resulting lens-induced anomalous parallax is
\begin{align}\label{eq:varpimag}
    \Delta \varpi_{li} =  -\frac{4 G M_l \varpi_l }{r_l \beta_l}  \widetilde{{\varpi}}_i(\beta_l, \vect{\beta}_{li}), 
\end{align}
where here $\vect{\beta}_{li} \simeq \vect{\beta}^0_{li}$ denotes to the zeroth-order impact parameter and the spatial profile is
\begin{align}\label{eq:varpiprofile}
	\widetilde{{\varpi}}_i(\beta_l, \vect{\beta}) & =  \frac{\widetilde{M}_l(\beta)}{\beta^2 / \beta_l^2}  \frac{(\cos\delta_\text{ecl})^2}{1+(\sin\delta_\text{ecl})^2} \left(2 \hat\beta^2_{\delta, \text{ecl}} - 1\right)  \nonumber \\
	&\phantom{=} + \frac{\partial_{\beta } \widetilde{M}_l(\beta )}{\beta /\beta_l^2} \frac{1 - (\cos\delta_\text{ecl})^2\hat{\beta}^2_{\delta, \text{ecl}} }{1 + (\sin\delta_\text{ecl})^2 }, 
\end{align}
where $\hat{\beta}_{\delta,\text{ecl}}$ denotes the component along the ecliptic latitude of the angular impact parameter unit vector. From the equation above, it is clear that the anomalous parallax effect is maximized at the ecliptic equator and suppressed at the poles. In analogy to the proper motion and angular acceleration templates, the parallax test statistic $\mathcal{T}_\varpi$ and its normalization $\mathcal{N}_\varpi$ can be defined as 
\begin{align} \label{eq:tau_par}
	\mathcal{T}_\varpi (\beta_{t}, \vect{\theta}_{t}) & = \sum_i \frac{\varpi_i  \widetilde{{\varpi}}_i(\beta_t, \vect{\beta}_{it})  }{\sigma_{\varpi, i}^2} \\
	\mathcal{N}_\varpi^2 (\beta_{t}, \vect{\theta}_{t} ) & = \sum_i \frac{ \widetilde{{\varpi}}^2_i(\beta_t, \vect{\beta}_{it})}{\sigma_{\varpi, i}^2},
\end{align}
where $\varpi_i$ is the measured parallax of the {\it i}-th star and $\sigma_{\varpi, i}$ the parallax dispersion.

Following the signal-to-noise ratio derivation of section~\ref{sec:vel_acc_proj}, we can forecast the sensitivity of the parallax template using
\begin{align} \label{eq:tau_par_snr}
    \SNR_{\mathcal{T}_\varpi} \simeq \frac{4 G M_l \AU}{r_l D_l} \frac{\sqrt{\Sigma_0}}{\sigma_{\varpi}},
\end{align}
where we have replaced $\varpi_l = \AU/D_l$ and assumed the maximum signal by taking $\sin\delta_\text{ecl}=0$, for simplicity. We can compare this with the SNR for the proper motion template
\begin{align} \label{eq:snr_pmpar}
    \frac{\SNR_{\mathcal{T}_\varpi}}{\SNR_{\mathcal{T}_\mu}} = \frac{\AU}{v_{l \perp}} \frac{\sigma_\mu}{\sigma_{\varpi}}.
\end{align}
When both the proper motion and the parallax uncertainties are statistically limited, $\sigma_\mu \propto \tau^{-3/2}$ and $\sigma_\varpi \propto \tau^{-1/2}$, and the above ratio becomes $\AU /v_{l \perp} \tau \sim 10^{-3}$, for a lens velocity of $v_{l \perp} \sim 10^{-3}c$ and a survey duration of $\tau = 10\ \textrm{yr}$. When instead the astrometric measurement can resolve the sources' intrinsic proper motion and parallax, $\sigma_\mu \simeq v_{i \perp} /D_i$ and $\sigma_\varpi \simeq \AU/D_i$, and the above ratio becomes $v_{i \perp}/v_l$, which is typically of $\mathcal{O}(1)$, unless we choose a stellar target with intrinsic velocity dispersion much smaller than the Milky Way's virial velocity.

We conclude by noting that the lens parallax motion induces additional corrections to the 2D stellar motion, beyond the one captured by the astrometric fit, given in equation (\ref{eq:varpimag}). These corrections -- as well as the angular acceleration of equation (\ref{eq:lens_acc}) -- are not modeled by a standard 5-parameter astrometric fit (like the one described in \cite{lindegren2012astrometric}, and used by the \gaia collaboration). Such an astrometric fit would then perform poorly on lensed luminous sources, leading to large values of the $\chi^2$. Unfortunately, there are several competing effects that can degrade the quality of the astrometric solution, and sources with a bad fit are not uncommon in the \gaia data \cite{2021A&A...649A...5F, 2021A&A...649A...2L}. It is therefore hard to rely uniquely on the $\chi^2$ observable to tease out lensing effects. A more effective strategy would be to perform a dedicated astrometric fit that includes additional parameters to model the lensing-induced acceleration and anomalous parallax corrections. This fit would presumably be computationally costly, but could be applied selectively on sources that are good candidates for a lensing event, i.e.~in locations where there is a large proper motion template and/or sources with a large $\chi^2$ from the 5-parameter fit.

\section{Accelerations from the DR2-DR3 position offset}
\label{app2:acceleration}

The astrometric solutions provided by the \gaia collaboration refer to a certain astrometric epoch, chosen so as to minimize the correlations between the position and proper motion fitting parameters \cite{brown2021gaia}. The reference epochs for DR2 and DR3 catalogs are J2015.5 and J2016.0, respectively, and this offset of 0.5 \text{y} has to be taken into account when comparing source positions between the two releases. 

Maps of the mean positional difference between the DR3 and DR2 \gaia catalogs -- having corrected for the difference in epoch -- reveal discrepancies that begin at the sub-\rm{mas} level fluctuating up to 2 \rm{mas}, and could be attributable to DR2 positional uncertainties \cite{2021A&A...649A...5F}. Since the astrometric solutions currently presented by the \gaia collaboration do not capture the full range of dynamics that may govern apparent stellar motions, these mismatches could also encapsulate accelerations, including of lens-induced, apparent accelerations in stars that have been fit only for parallax, position, and proper motion.

Reconstructing angular accelerations for stars by combining their astrometric measurements at different epochs was made possible through the cross-calibration of the {\it Hipparcos} and \gaia catalogs in~\cite{brandt2018hipparcos, brandt2021hipparcos}. 
The clear advantage in combining observations from these two missions is the long baseline of $\sim 24$ \text{y}, while the drawback is the limited number of available sources, just over $115,000$. 

Lacking access to \gaia's transit timing information for each source, which would allow us to fit for nonlinear trajectories and look for lens-induced accelerations, we present here the derivation of an alternative estimator for calculating stellar accelerations \{$\vect{\alpha}_{i}$\}, by leveraging the difference in epoch between different data releases. 
By cross-matching the DR2 and DR3 \gaia datasets, we can leverage the larger number of matched sources and create a catalog that can serve as a look-up table for stellar accelerations, albeit potentially being limited by the shorter observational baseline of $\sim 0.5$ y between the two releases. A machine-learning algorithm that takes advantage of this difference in \gaia's observational timelines has generated a catalog of nearby accelerating star candidates at high statistical significance \cite{2023AJ....165..193W}. We propose a complementary analytic approach that can be used on all sources with a 5-parameter astrometric solution.

We simplify \gaia's astrometric fit by only considering the stars' 2D motion in the plane perpendicular to the line of sight. We further assume uncorrelated uncertainties $\sigma_{\delta \theta}$ per individual observation in both angular directions, and that the position of each source is recorded at regular time intervals with frequency~$f$. Furthermore, each source is observed for a total time $\tau$ and \gaia's best-fit parameters $\lbrace \bar{\theta}_{\textit{G}}, \bar{\mu}_{\textit{G}}, \bar{\varpi}_{\textit{G}} \rbrace$ are given at the midpoint of the observations. The astrometric fit will be the result of minimizing the following test statistic
\begin{equation}
\chi^{2} = \sum \limits_{i=0}^{f\tau}  \frac{[\theta(t_i)- \theta_G (\bar{\theta}_{\textit{G}}, \bar{\mu}_{\textit{G}}, \bar{\varpi}_{\textit{G}} |t_{i} )]^{2}}{\sigma_{\delta \theta}^{2}},
\label{eq:chi2}
\end{equation}
where the stellar motion in either direction is modeled as
\begin{equation}
\theta_G (\bar{\theta}_{\textit{G}}, \bar{\mu}_{\textit{G}}, \bar{\varpi}_{\textit{G}} | t )= \bar{\theta}_{\textit{G}} + \bar{\mu}_{\textit{G}}\left(t - \frac{\tau}{2}\right) + \bar{\varpi}_{\textit{G}} \cos\left[\frac{2\pi}{T} \left(t-\frac{\tau}{2}\right) \right],
\label{eq:gaia_fit}
\end{equation}
and its true motion, in the presence of an angular acceleration $\bar{\alpha}$, is 
\begin{align}
    \theta(t)  & = \bar{\theta} + \bar{\mu}\ t  + \frac{\bar{\alpha}}{2}t^2 + \bar{\varpi} \cos\left(\frac{2\pi}{T} t\right) ,
\end{align}
where $T=1\ \rm{y}$. For sufficiently frequent observations over more than one year, we can integrate Eq.~\eqref{eq:chi2} over the total observation time $\tau$ and find the best-fit solution by minimizing the $\chi^2$ with respect to \gaia's parameters. The solution allows us to write the true stellar motion parameters in terms of $\lbrace \bar{\theta}_{\textit{G}}, \bar{\mu}_{\textit{G}}, \bar{\varpi}_{\textit{G}} \rbrace$ and $\bar{\alpha}$ as
\begin{align} 
    \label{eq:thetacomp}
    &\bar{\theta} =  \bar{\theta}_{\textit{G}} - \bar{\mu}_{\textit{G}} \frac{\tau}{2} + \frac{\bar{\alpha}}{12} \tau^2 \\ 
    \label{eq:mucomp} 
    &\bar{\mu} =  \bar{\mu}_{\textit{G}} - \frac{\bar{\alpha}}{2} \tau. \\
    \label{eq:varpicomp} 
    &\bar{\varpi} =  \bar{\varpi}_{\textit{G}} - \frac{\alpha T^2}{2\pi^2},
\end{align}
where the expressions have been further simplified by assuming that $\tau$ is an integer multiple of $T$. Due to the acceleration term, astrometric fits performed over different observational periods -- and therefore at different reference epochs -- will not give the same result for $\bar{\theta}_{\textit{G}}$ and $\bar{\mu}_{\textit{G}}$. Since the left-hand-side of the equations above is fixed, the acceleration can be expressed in terms of the DR2-DR3 positional offset $\Delta \bar{\theta}_{23} = \bar{\theta}_{G_2} - \bar{\theta}_{G_3}$ and the measured proper motions as 
\begin{align}
\bar{\alpha} =  \frac{12}{\tau_{3}^2 -\tau_{2}^2 } \left(\Delta \bar{\theta}_{23}  + \frac{\bar{\mu}_{G_3}\tau_{3} - \bar{\mu}_{G_2}\tau_{2}}{2} \right).
\label{eq:acc1}
\end{align}
Similarly, the acceleration could also be obtained from Eq.~\eqref{eq:mucomp} in terms of the proper motion offset $\Delta \bar{\mu}_{23} = \bar{\mu}_{G_2} - \bar{\mu}_{G_3}$, 
\begin{align}
\bar{\alpha} =  -2 \frac{\Delta \bar{\mu}_{23} }{ \tau_{3}-\tau_{2} }.
\label{eq:acc_mu}
\end{align}
By doing error propagation in equation (\ref{eq:acc1}), we get 
\begin{equation}
\sigma_{\alpha_{\Delta \theta}}= \frac{12}{(\tau_{3}^{2} - \tau_{2}^{2})} \sqrt{\sigma_{\theta_{2}}^{2} + \sigma_{\theta_{3}}^{2} + \frac{\tau_{3}^{2}\sigma_{ \mu_{3}}^{2}}{4} + \frac{\tau_{2}^{2}\sigma_{ \mu_{2}}^{2}}{4}  },
\label{eq:offsetunc1}
\end{equation}
where $ \sigma_{ \theta_{i}}$ ($ \sigma_{ \mu_{i}}$) denotes the angular position (proper motion) uncertainty of data set $i$. The errors on the best-fit parameters of the astrometric solution take the following form \citepalias{2018JCAP...07..041V} 
\begin{equation} 
\sigma_{\theta} = \frac{3\sigma_{\delta \theta}}{2\sqrt{f \tau}}; \quad
\sigma_{\mu} = \frac{2 \sqrt{3}\sigma_{\delta \theta}}{\sqrt{f \tau^{3}}}
\label{eq:inst_precision},
\end{equation}
which leads to,
\begin{align}
\sigma_{\alpha_{\Delta \theta}} & = \frac{6\sqrt{21} }{\tau_3 - \tau_2} \frac{\sigma_{\delta \theta}}{\sqrt{f \tau_2\tau_3(\tau_2+\tau_3)}} \simeq 300\ \mu\rm{as}/\rm{y}^2 \left(\frac{\sigma_{\delta \theta}}{200\ \mu\rm{as}}\right),
\label{eq:offsetunc2}
\end{align}
for $\tau_{2(3)} = 22\ (34)\ \rm{months}$ and $f=14\ \rm{y}^{-1}$. If the angular acceleration were instead measured by performing a 7-parameter astrometric fit for each source, the expected uncertainty would be given, in analogy to equation (\ref{eq:inst_precision}), by 
\begin{equation}
\sigma_{\alpha_7}
\label{eq:a7} = \frac{12 \sqrt{5} \sigma_{\delta \theta}}{\sqrt{f \tau^{5}}} \sim 20\ \mu\rm{as}/\rm{y}^2 \left(\frac{\sigma_{\delta \theta}}{200\ \mu\rm{as}}\right) \left(\frac{66\ \rm{months}}{\tau} \right)^{5/2}.
\end{equation}
The positional offset approach is worse than an acceleration-based analysis by a factor of
\begin{align}
\frac{\sigma_{\alpha_{\Delta \theta}}}{\sigma_{\alpha_7}} & = \frac{1}{2(\tau_3 - \tau_2)} \sqrt{\frac{21}{5} \frac{\tau^5}{\tau_2\tau_3(\tau_2+\tau_3)}}
    	& \simeq  \begin{cases}
     		 15  \qquad \tau = 5.5\ \text{y} \\
     		 66  \qquad \tau = 10\ \text{y}
   	 \end{cases}\,,
\label{eq:ratio}
\end{align}
where we forecast the results for a 7-parameter fit performed by using time-series observations for either the DR4 catalog, or the final data release of the \gaia mission. Nevertheless, our method can be applied on currently available data and does not present the computational challenges of a higher-parameter fit on the large number of \gaia sources. As shown in Fig.~\ref{fig:template_projections}, an acceleration template analysis could already produce interesting results in the search for lensing signals from dark compact lenses. We leave the construction of a full DR2/DR3 acceleration catalog, including the study of potential systematic uncertainties, and the development of the corresponding template analysis to future work~\cite{acc_paper}.

%%%%%%%%%%%%%%%%%%%%%%%%%%%%%%%%%%%%%%%%%%%%%%%%%%

	% Don't change these lines
	\bsp	% typesetting comment
	\label{lastpage}
\end{document}